\documentclass[fleqn,usenatbib,useAMS]{mnras}

\usepackage{graphicx}	
\usepackage{amsmath}	
\usepackage{multicol}        
\usepackage{bm}		
\usepackage{pdflscape}	
\usepackage{siunitx}
\usepackage{booktabs}

\usepackage[T1]{fontenc}
\usepackage{ae,aecompl}

\usepackage{newtxtext,newtxmath}

\title[$\alpha$--Fe bimodality in constrained simulations of the Local Group]{Understanding the formation of $\alpha$--Fe bimodality in the Milky Way using constrained cosmological simulations of the Local Group}

\author[A. Genina et al.]{Anna Genina$^{1,2}$%
\thanks{Contact e-mail: \href{mailto:agenina@mpa-garching.mpg.de}{agenina@ed.ac.uk}}, %
Ewoud Wempe$^{3,4}$,
Amina Helmi$^{3}$,
Hanneke C. Woudenberg$^{3}$, Simon D. M. White$^{2}$,
\newauthor
Robert J. J. Grand$^{5}$,
R\"udiger Pakmor$^{2}$, Guilhem Lavaux$^{6}$, Jens Jasche$^{7}$
\\
$^{1}$Institute for Astronomy, University of Edinburgh, Royal Observatory, Blackford Hill, Edinburgh EH9 3HJ, UK \\
$^{2}$Max Planck Institute for Astrophysics, Garching bei M\"unchen, 85748, Germany \\
$^{3}$Kapteyn Astronomical Institute, University of Groningen, P.O.Box 880, 9700 AV Groningen, The Netherlands \\
$^{4}$LIRA, Observatoire de Paris, Université PSL, Sorbonne Université, Université Paris Cité, CY Cergy Paris Université, CNRS, 92190 Meudon, France \\
$^{5}$Astrophysics Research Institute, Liverpool John Moores University, 146 Brownlow Hill, Liverpool, L3 5RF, UK \\
$^{6}$CNRS \& Sorbonne Université, UMR 7095, Institut d’Astrophysique de Paris, 98 bis boulevard Arago, F-75014 Paris, France \\
$^{7}$The Oskar Klein Centre, Department of Physics, Stockholm University, Albanova University Center, SE 106 91 Stockholm, Sweden
}
\date{Last updated 2024 April 23; in original form 2013 September 5}

\pubyear{{\the\year{}}}

\begin{document}
\label{firstpage}
\pagerange{\pageref{firstpage}--\pageref{lastpage}}
\maketitle

\begin{abstract}

We use a novel suite of constrained cosmological simulations of the Local Group to investigate the formation of $\alpha$--Fe bimodalities in stellar discs of Milky Way analogues. In agreement with past studies using the {\sc Auriga} model, we show that in our simulations the high-$\alpha$ sequence forms in a merger-induced starburst, which further quenches star formation, allowing the new stellar populations to form at low [$\alpha$/Fe]. Producing a bimodality that is prominent at large radii, with a significant [$\alpha$/Fe] gap and overlapping metallicity as in the Milky Way, requires the quenching of star formation across all radii where the bimodality is observed. The quenching can be achieved either with a galaxy-wide starburst, or with a retrograde merger reducing angular momentum in the outer disc and allowing more efficient dilution by the metal-poor gas in the circumgalactic medium (CGM). We show that mergers can also tilt the disc, changing the direction of galactic outflows and enriching a larger volume of the CGM with galactic metals, leading to low [$\alpha$/Fe] scatter in the subsequently formed low-$\alpha$ sequence. We argue that the slopes in [$\alpha$/Fe] of the two $\alpha$ sequences are crucial in narrowing down the mechanism that leads to their formation. We additionally uncover an indirect role that mergers play in enhancing the accretion of metal-poor gas onto galaxies. The smoothly accreted gas can dominate the gas content in the CGM and aid in the formation of distinct chemical sequences, especially at the low-metallicity end.

\end{abstract}





\section{Introduction}

In the Cold Dark Matter (CDM) model, galaxies like the Milky Way form inside more massive dark matter haloes in which gas can shock-heat and radiatively cool onto a disc \citep{white_rees,Fall:1980sj,Blumenthal:1984bp} or possibly directly fuel the disc via cold flows at high-redshift (\citet{fardal,keres,  dekel_shock_cool,ocvirk, Brooks2009, Danovich2012, Danovich2015, Rivers2021}; however see \citet{nelson_sph_arepo, nelson2016zooming} for caveats).
Although dark matter haloes are built hierarchically in the CDM model, the role of mergers is only secondary in the formation of galactic discs.

Chemo-dynamical studies based on surveys like \textit{Gaia} \citep{gaia_dr2, gaia_edr3}, APOGEE \citep{apogee} and GALAH \citep{galah} have been revealing a complex picture regarding the formation and structure of the Milky Way. One particular peculiarity of the Milky Way's stars that has received significant attention is the bimodality in the $\alpha$-element -- metallicity plane, observed in the Solar neighbourhood \citep{bensby, fuhrmann}. The high and low-$\alpha$ populations appear to correlate kinematically with the thick and thin disc, respectively \citep{haywood_two_phase}. The high-$\alpha$ population consists mainly of old stars on kinematically hot orbits with some amount of rotational support, while the low$-\alpha$ population is kinematically cold and consists mostly of younger stars \citep{adibekya,recio-blanco,hayden,queiroz_starhorse}, although recent studies by \citet{nepal}, \citet{gent} and \citet{barbolato} suggest that the two populations did not form sequentially but were coevolving. Other studies have also found overlap in age of the stars in the low and high-$\alpha$ sequences \citep{hayden_thick_or_thin, mackreth}. Interestingly, the majority of the low-$\alpha$ population appears to have formed in the past 10~Gyr and high-$\alpha$ population relatively quickly between 9 and 12~Gyr ago \citep{snaith_thick_disk, silva_aguirre}. This 10~Gyr threshold corresponds to the inferred arrival of the Gaia Enceladus/Sausage \citep{gallart}, which is believed to be the last significant merger our Galaxy has experienced \citep{enceladus,sausage,haywood_lmm}. 

Several hypotheses have been proposed to explain the origin of the two-$\alpha$ sequences. In the ``two-infall'' model \citep{chiappini1,chiappini2}, the high-$\alpha$ sequence forms during an early phase of rapid gas accretion and star formation, followed by a later, more extended accretion episode that forms the low-$\alpha$ disc from lower-metallicity gas. The time delay between these episodes ensures a decline in [$\alpha$/Fe], while metal dilution in the second episode of gas infall produces the overlap in metallicity between the two sequences. ``Parallel'' models, in which the thin and thick discs form at the same time but with different rates \citep{grisoni2017}, have also been proposed; however, reproducing the age distribution of old stars appears to favour the two-infall model, with a separation between the two infall episodes of approximately 3--5~Gyr \citep{spitoni2019,spitoni_2021_apogee,spitoni2024}. The intense star formation in the first infall episode could be related to the Gaia Enceladus/Sausage (GES) merger or to the establishment of the Galactic bar \citep{haywood_2018}.

Cosmological simulations provide a way to connect these chemical evolution models to realistic assembly histories of Milky Way-like galaxies within CDM. A common trend emerging from several studies is that the high-$\alpha$ sequence requires an early phase of efficient star formation and rapid gas consumption, followed by a reduction in the star formation rate, with the low-$\alpha$ sequence forming quiescently from lower-metallicity gas. In some simulations, this transition is driven by a gas-rich merger, which can both trigger intense star formation and supply low-metallicity gas \citep{grand_discs, buck_2020, beane1}. In others, the main ingredient is not the merger itself, but the subsequent infall of metal-poor gas from the circumgalactic medium, which dilutes the star-forming gas and allows the low-$\alpha$ sequence to form at low metallicity \citep{fire_bimodality, orkney}. These results suggest that mergers may trigger or facilitate the formation of the low-$\alpha$ sequence, but that they are not necessarily the only source of the metal-poor gas required in its formation. 

The importance of mergers in the formation of a chemical bimodality or the prevalence of such bimodalities in Milky Way-like galaxies remains uncertain. Using NIHAO-UHD simulations, \citet{buck_2020} find that a gas-rich minor merger can naturally produce a chemical bimodality by diluting the metallicity of the star-forming gas after the high-$\alpha$ sequence is in place (see also \citealt{grand_discs}). They argue this process is likely not uncommon in disc galaxies. By contrast, \citet{mackereth_2018} find in {\sc EAGLE} simulations that Milky Way-like bimodality is rare and requires unusually rapid early mass growth followed by later gas infall, suggesting that the Milky Way may have had an atypical formation history. {\sc Auriga} and FIRE-2 studies similarly find that bimodality is not universal, but is most strongly associated with early intense star formation followed by a quiescent phase \citep{grand_discs,fire_bimodality,orkney,barry_fire}, often triggered by a merger; however, \citet{amarante} and \citet{johnson_retro} argue that only a retrograde GES-like merger can produce a strong bimodality through a central starburst and the reduction of star formation in the outer disc of the Milky Way.  

Mergers are not the only way to induce a burst in star formation with a subsequent reduction in the star formation rate. Similar star formation histories can also be achieved for example through strong ejective feedback at early times \citep{khoperskov}, or through a transition from cold to hot-mode accretion with delayed enrichment by Type Ia supernovae \citep{noguchi, funakoshi}. \citet{beane2} also argue that bar formation and Active Galactic Nuclei (AGN) can induce the necessary gap in star formation. \citet{grand_discs} show that the exhaustion of gas in the outer disc (``disc shrinking'') can also provide the necessary gap in star formation to create a chemical bimodality. 

Alternative mechanisms for the formation of the [$\alpha$/Fe] bimodality have also been proposed that do not require mergers or a hiatus in star formation. Radial migration of high-$\alpha$ stars formed in the thick disc was initially believed to explain bimodality at the Solar radius \citep{schroenrich, loebman,sharma,chen}, but this scenario struggles to explain the bimodality across the disc (\citealt{imig}, see also \citealt{minchev1,minchev2}). Fragmentation of the disc at high redshift into clumps that form higher-$\alpha$ stars compared to the smooth and slower evolving disc could also explain the dichotomy \citep{clarke, beraldo_e_silva} and would imply coevolution of the thin and thick discs as inferred by \citet{barbolato}. Using VINTERGATAN simulations, \citet{agertz} and \citet{vintergatan3} propose a ``tilting-disc'' scenario, in which a gas-rich satellite feeds the formation of a misaligned, low-$\alpha$ metal-poor disc that subsequently aligns with the pre-existing high-$\alpha$ disc, with the rest of the low-$\alpha$ population forming secularly over the following several Gyrs. 

Overall, the formation of the Milky Way's two chemical sequences is likely the result of several processes, some of which may be linked: early rapid star formation, gas accretion and dilution, gas metal enrichment and galactic fountain flows, radial migration, and likely one or more significant mergers. In this work, we study the formation of chemical sequences in the discs of Milky Way-like galaxies using constrained cosmological simulations aimed at reproducing the key properties of the Local Group. While the large-scale structure is not directly constrained in our simulations, the Milky Way-M31 pairs are embedded in a sheet-like structure as a result of the constraints imposed using the motions of galaxies out to 4 Mpc  \citep{wempe_sheet}. Moreover, one of our Milky Way analogues follows a history comparable to that inferred for the Milky Way, including a GES-like merger, an extended interaction with a Sagittarius-like satellite and the arrival of a Large and Small Magellanic Cloud-like pair, making it particularly interesting in the context of the thin and thick disc formation in our Galaxy. By construction, our simulations contain M31 as well as the Milky Way; however, here we focus on investigating the [$\alpha$/Fe] bimodality in the Milky Way, where its existence is significantly more evident than it is in the M31 \citep{kobayashi_m31, gibson_m31}, although we aim to address the chemical properties of M31's disc in future work.  

This paper is structured as follows: in Section~\ref{sec:sims} we describe the initial conditions, cosmology and the galaxy formation model used in our simulations, as well as present the structural properties of our Milky Way analogues. In Section~\ref{sec:galaxy_assembly} we study the mass assembly of the simulated galaxies, focusing on gas infall during significant galaxy mergers. In Section~\ref{sec_form_bim} we study the presence of chemical bimodalities in the simulated systems and their origin. Section~\ref{sec:discussion} discusses our findings in the context of the Milky Way's chemical bimodality and its properties in comparison to our analogues, and in Section~\ref{sec:conclusions} we conclude.

\section{Simulations overview}
\label{sec:sims}

\begin{figure*}
    \includegraphics[width=0.5\columnwidth]{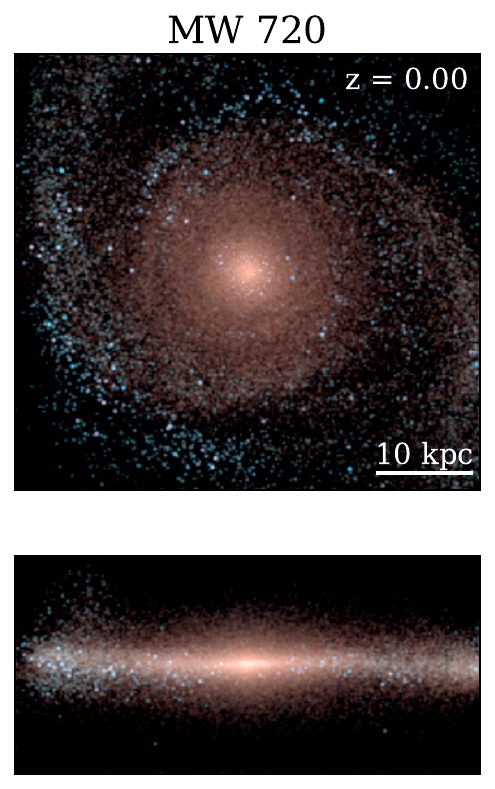}
    \includegraphics[width=0.5\columnwidth]{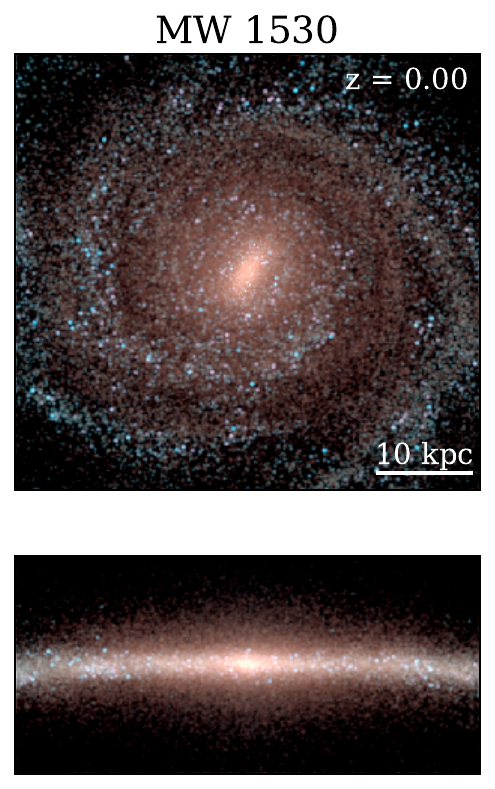}
    \includegraphics[width=0.5\columnwidth]{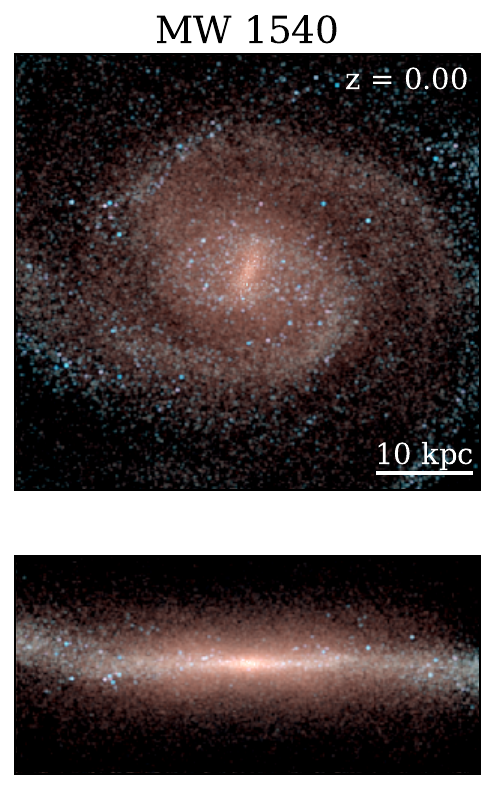}
    \includegraphics[width=0.5\columnwidth]{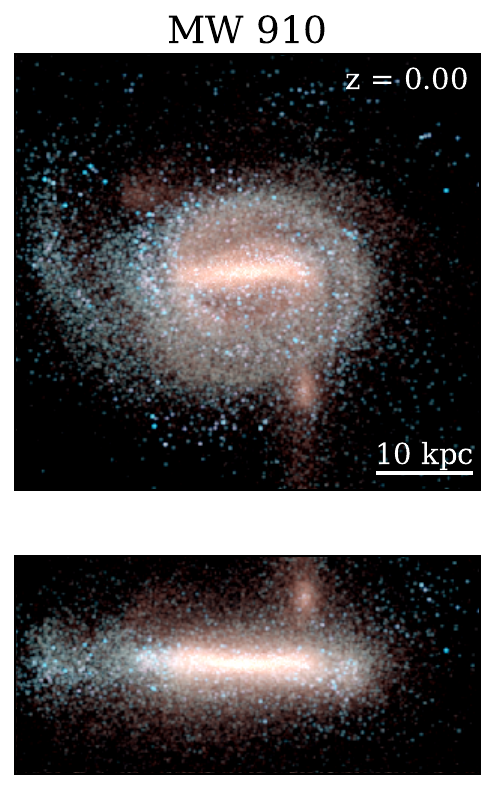}
    \caption{\textit{Top:} face-on projections of our 4 simulated Milky Ways at $z$=0. The stellar colours are computed using the \textit{U}, \textit{B} and \textit{K} bands \citep{bruzual_charlot} and include all stars, \textit{in-situ} and accreted within the 50~kpc box centred on the galaxy. \textit{Bottom:} edge-on projections. Vertically offset from the disc of MW 910 we see what may be considered an analogue of the Small Magellanic Cloud (because of its infall time and mass).}
    
    \label{fig:present-day}
\end{figure*}

\subsection{Initial conditions}

In this work, we use initial condition (IC) fields from the constrained cosmological simulations of \citet{Wempe2024}, produced using Bayesian field-level inference \citep{jasche_wandelt, lavaux_jasche, jasche_lavaux}. These are a fair and representative sample of $\Lambda$CDM initial conditions, and are conditioned\footnote{We note that neither the merger histories of the Milky Way analogues, nor the subsequent stellar mass growth were constrained in the creation of initial conditions. The emergence of galactic discs in the Milky Way analogues and the presence of chemical bimodalities within any of them are therefore the predictions of the galaxy formation model, given the initial conditions.} to match the present-day masses, positions, and relative velocity of the MW-M31 pair, as well as the surrounding flow field within \SI{\sim4}{Mpc}. Of these, 224 realisations were resimulated at higher resolution in dark-matter-only by \citet{Wempe2025}; here, we select four to run with full physics. The first three (C1 S720, C6 S1530, C3 S1540, hereafter 720, 1530 and 1540) were chosen randomly, and the fourth (C10 S910, hereafter 910) was selected because its MW analogue has also Large Magellanic Cloud (LMC), Small Magellanic Cloud (SMC), GES and Sagittarius analogues \citep{Wempe2025}. The dark matter particle initial conditions were generated at $z=63$, assuming $\Lambda$CDM cosmology, with $H_0=67.556$~km~s$^{-1}$~Mpc$^{-1}$, $\Omega_0 = 0.312$, $\Omega_{\Lambda} = 0.688$ and $\Omega_b = 0.048$.

The initial conditions for the magnetohydrodynamics (MHD) simulations were generated by splitting all dark matter particles in the dark-matter-only ICs into a dark matter particle and a hydrodynamics mesh-generating point, following the approach adopted in the {\sc Auriga} \citep{grand_auriga,grand_auriga_new} and {\sc Illustris-TNG} \citep{pillepich_tng} MHD simulations with {\sc Arepo} \citep{arepo}. The original initial conditions of \citet{Wempe2024} consist of a 40~Mpc$^3$ low-resolution region and a 10~Mpc$^3$ high-resolution region. The high-resolution region therefore extends only marginally beyond the Local Group. The enforced cubic shape of the high-resolution region and a significant difference between the low and high-resolution particle masses can result in the migration of low-resolution particles to the high-resolution region. To mitigate numerical effects stemming from low-resolution particles and cells entering the high-resolution region, we increase the number of particles in the low-resolution region to 128$^3$ particles, compared to the original 64$^3$ of \citet{Wempe2024}, with a 256$^3$ dark matter particles in the high-resolution region.

\subsection{Galaxy formation model}

Our simulations employ the {\sc Auriga} galaxy formation model\footnote{We have also tested our conclusions using the \textsc{TNG} galaxy formation model and obtained qualitatively similar results regarding the shape of the [Mg/Fe]-[Fe/H] distribution in our Milky Way analogues and the origin of the chemical sequences.}, which is described in detail in \citet{grand_auriga} and is implemented in the {\sc Arepo} moving-mesh MHD code \citep{arepo}. In brief, the reionisation is modelled as a homogeneous field and completes at $z\approx6$, following \citet{faucher_giguere}. Primordial and metal-line cooling with a correction for self-shielding are modelled as in \citet{vogelsberger2013}.  Magnetic fields are implemented as in \citet{mag_fields1} and \citet{magnetic_fields}. The interstellar medium (ISM) gas follows the equation of state of \citet{sh03}, with star formation occurring in gas at densities above $n > 0.13 \rm cm^{-3}$. Once the gas reaches the threshold density, the star formation rate is computed, which is used to determine the probability of forming a star particle. The star-forming gas cells can also be transformed into ``wind'' particles. For each star-forming gas cell the wind velocity and energy loading are computed, which determine the mass loading and thus the probability for a star-forming cell to become a wind particle, as opposed to a star particle. Star particles represent a simple stellar population following a \citet{chabrier} initial mass function. If a wind particle is formed, it is ejected at a velocity proportional to halo velocity dispersion and is decoupled from the hydrodynamics. Once either the maximum travel time (given as 0.025 of the Hubble time at the time of ejection) has elapsed or the wind particle reaches a local density equal to 0.05 of the star formation density threshold, it is recoupled with the gas, injecting 50~per~cent thermal and 50~per~cent kinetic energy \citep{marinacci_2014}. The wind particle also deposits its mass and metals into the gas cell. Enrichment from Asymptotic Giant Branch stars is computed with yields from \citet{karakas}, type II supernovae from \citet{portinari} and type Ia supernovae from \citet{thielmann} and \citet{travaglio}.

Black holes of $M_{\rm BH} \!=\!10^5 M_{\odot} h^{-1}$ are seeded in haloes once they reach the halo mass $M_{\rm h}=5\times10^{10} M_{\odot} h^{-1}$. Black hole accretion and mergers follow the prescription of \citet{springel_bh} and quasar and radio mode feedback are modelled with thermal energy injection in the vicinity of the black hole and in the halo (see \citealt{grand_auriga}).

As our simulations are closest in resolution to Auriga level-5, with dark matter particle mass of $m_{\rm DM}\!=\!2\times10^6 M_{\odot}$ and $m_{\rm gas}\!=\!3.6\times10^5 M_{\odot}$, we likewise use the minimum physical softening of $\epsilon_{\rm phys} \!=\! 0.74$~kpc for stars and dark matter in the high-resolution region.

\begin{table*}
\centering
\caption[]{Properties of our four MW analogues,  columns showing (1) their IDs, (2) virial mass, excluding the mass in satellites, (3) virial radius, (4) stellar mass within $R_{200}$, (5) disc mass including thin and thick discs, (6) mass in the spheroidal component within 0.1$R_{200}$, (7) mass of the bar, (8) the ratio of the thick and thin disc mass, (9) the bulge-to-total ratio,$(M_s + M_{\rm bar})/(M_s + M_{\rm bar} + M_{d})$, (10) fraction of accreted stars within 0.1R$_{200}$, (11) fraction of gas within 0.1$R_{200}$. Values in brackets account for the mass of individual components inside the region occupied by the bar (see Appendix~\ref{disk_params}). Signed values show the additional mass and unsigned values show recomputed values with mass in the region of the bar accounted for. }
\begin{tabular}{lcccccccccc}
\toprule
\textbf{Run} &
\multicolumn{1}{c}{$\dfrac{M_{200}}{10^{10} \mathrm{M}_\odot}$} &
\multicolumn{1}{c}{$\dfrac{R_{200}}{\rm kpc}$} &
\multicolumn{1}{c}{$\dfrac{M_*}{10^{10} \mathrm{M}_\odot}$} &
\multicolumn{1}{c}{$\dfrac{M_d}{10^{10} \mathrm{M}_\odot}$} &
\multicolumn{1}{c}{$\dfrac{M_s(<0.1 R_{200})}{10^{10} \mathrm{M}_\odot}$} &
\multicolumn{1}{c}{$\dfrac{M_{\rm bar}}{10^{10} \mathrm{M}_\odot}$} &
${\dfrac{\rm Thick}{\rm Thin}}$ &
${\dfrac{\rm B}{\rm T}}$ &
$f_{\mathrm{acc}}$ &
$f_{\mathrm{gas}}$ \\
\midrule
MW 720 & 94.16   & 207.06 & 4.88 & 3.61 & 0.87 & 0.00 & 2.17 & 0.20  & 0.28 & 0.07 \\
MW 1530 & 152.41 & 243.19 & 7.20 & 4.12 (+0.52) & 0.93 (+0.45) & 1.70 (0.73) & 1.04 (1.17) & 0.39 (0.31)  & 0.09 & 0.08 \\
MW 1540 & 110.41 & 218.38 & 5.54 & 4.23 (+0.34) & 0.60 (+0.31) & 0.58 (--) & 1.43 (1.56) & 0.22 (0.16)  & 0.13 & 0.08 \\
MW 910 & 130.02  & 230.69 & 8.33 & 2.99 (+0.93) & 1.40 (+0.40) & 3.67 (2.34) & 1.03 (0.95) & 0.63 (0.51)  & 0.13 & 0.12 \\
\bottomrule
\end{tabular}
\label{tab:main-props}
\end{table*}

\begin{table*}
\centering{}
\caption[]{The disc and bar properties of the four MW analogues identified in the first column, columns showing the (2) $r$-band half-light radius for stars within 17~kpc, (3) scale radius of the thin disc, (4) scale height of the thin disc, (5) scale radius of the thick disc, (6) scale height of the thick disc, (7) bar strength and (8) bar length. It is worth noting that the stiff \citet{sh03} equation of state in the {\sc Auriga} model increases the pressure support and stability of the gas discs, effectively setting a minimum scale height that can be produced in the simulations, exceeding that of the Milky Way's thin disc at this resolution \citep{burger}. 
}
\begin{tabular}{lccccccc}
\toprule
\textbf{Run} &
\multicolumn{1}{c}{$\dfrac{R_{\rm h}}{\rm kpc}$} &
\multicolumn{1}{c}{$\dfrac{R_{\rm thin}}{\rm kpc}$} &
\multicolumn{1}{c}{$\dfrac{h_{z,\rm thin}}{\rm kpc}$ } &
\multicolumn{1}{c}{$\dfrac{R_{\rm thick}}{\rm kpc}$} &
\multicolumn{1}{c}{$\dfrac{h_{z,\rm thick}}{\rm kpc}$ } &
$A_{2,{\rm max}}$ &
\multicolumn{1}{c}{$\dfrac{l_{\rm bar}}{\rm kpc}$ } \\
\midrule
MW 720 & 7.9 & 9.1 & 0.7 & 3.6 & 1.7 & -- & --\\
MW 1530 & 7.3 & 7.1 & 0.6 & 3.4 & 1.8 &  0.3 &  3.6 \\
MW 1540 & 8.6 & 7.6 & 1.2 & 3.2 & 2.5 & 0.2 & 2.6 \\
MW 910 & 5.8 & 3.5 & 1.2 & 3.6 & 1.9 & 0.6 & 6.3 \\
\bottomrule
\end{tabular}
\label{tab:main-props2}
\end{table*}

Since {\sc Arepo} propagates the gas-cells in a pseudo-Lagrangian fashion and the cells are permitted to split and merge, we use tracer particles to track the evolution of each gas parcel using the method of \citet{genel}. Each gas cell in the initial conditions is loaded with 8 tracer particles (permitted by the relatively low resolution we have here), allowing us to track the flow of gas down to star formation as well as ejection of wind particles and black hole accretion and feedback.

\subsection{Merger trees and in-situ stars}

For each simulation, we obtain 337 time outputs between $z=15$ and $z=0$, equally spaced in scale factor, allowing us to examine the full accretion and merger histories in detail. The merger trees and halo catalogues were constructed using {\sc SUBFIND-HBT}, developed as part of the {\sc Gadget-4} code \citep{gadget4}, which utilizes the {\sc HBT +} algorithm \citep{hbt}, allowing more faithful identification of the halo main progenitor by following the most bound particles at each snapshot . Moreover, the algorithm is advantageous for close pericentric passages of subhaloes or mergers, preventing misclassification of subhalo particles as the main halo. Despite these advantages, in some cases the algorithm may still misidentify some of the particles of the merger with those of the main halo. In those cases, we remove from our analysis the snapshots (typically no more than one) where the misidentification occurs.

The above procedure is essential for the identification of in-situ and ex-situ stars, gas and dark matter that we use in the subsequent analysis. In particular, to select in-situ stars, we identify the star particles in the Milky Way analogue haloes at $z=0$ that have never been bound to a halo or subhalo that is not the Milky Way main progenitor. For dark matter, we first select all particles that were ever bound to the Milky Way analogue. If any of these particles were at any point bound to haloes other than the Milky Way analogue, they are classified as having been contributed by mergers. The remainder of the particles are classified as ``smoothly'' accreted, although we note that some of these may represent dark matter haloes below our resolution limit of $M_{\rm halo}\sim6.4\times10^{7}$ M$_{\odot}$ (resolved with 32 particles). The classification of gas as coming from mergers or via smooth accretion is carried out by tracking past halo associations of tracer particles. The ``smoothly'' accreted gas constitutes the tracer particles that have never belonged to gas cells bound to haloes other than the Milky Way analogues, although they could have been at one point expelled from a Milky Way analogue and later reaccreted.

\subsection{Properties of the Milky Way analogues}

Fig.~\ref{fig:present-day} shows the present-day face-on (top panels) and edge-on (bottom panels) distribution of stellar particles in the simulated Milky Ways, with colours shown in the \textit{U}, \textit{B} and \textit{K} bands. The photometry has been computed using the tables of \cite{bruzual_charlot} on-the-fly in the simulation. In each case, the Milky Way-analogues form disc-dominated galaxies, in some cases showing prominent spiral arms (MW 1530) and a strong bar (MW 910).

The detailed galaxy properties\footnote{Throughout this work, we use the definition of the virial mass, $M_{200}$ as the mass within the radius $R_{200}$, enclosing the density equal to 200 times the critical density of the Universe.} are shown in Tables~\ref{tab:main-props} and~\ref{tab:main-props2}. By construction, the halo masses are within the range of previous dynamical estimates \citep{wang_milkyway}, while stellar masses are in some cases on par with observational estimates (MW 1530, MW 1540), and in others slightly below (MW 720) or above (MW 910) \citep[see for reference, e.g.][]{milky_way_stellar}.

To derive the contributions of the thin and thick discs, the bar, as well as the mass contribution of the spheroidal components, we first identify stellar particles belonging to the bar and perform a decomposition on the remainder of the particles based on their orbital circularity \citep{abadi_circularity}. The decomposition allows for two disc components and up to six spheroidal components\footnote{We limit the disc decomposition to two components; however we note that the star formation histories and chemical abundances of some of our analogues suggest multiple chemo-kinematic disc components.}. Since the removal of bar particles also removes some of the contribution of the discs and the spheroids from the centres of the galaxies, we additionally estimate the mass contributed by these components in the region occupied by the selected bar particles. These are shown in brackets in Table~\ref{tab:main-props} together with the revised bar mass, thick-to-thin disc ratios and the bulge-to-total ratios. The full methodology is detailed in Appendix~\ref{disk_params}.

The disc structural parameters were obtained by fitting a double-exponential disc profile to the surface density of the circularity-selected thick and thin discs. The resulting disc exponential scale length and heights are shown in Table~\ref{tab:main-props2}, along with the bar parameters. Most galaxies exhibit a spatially extended thin and a more concentrated thick disc, aside from MW 910 which has comparable scale lengths for the two components, which however differ in scale height. This galaxy also has a bar strength of 0.6 -- the only one in the sample which can be classified as a strong bar. This is in agreement with the findings of \citet{frakgoudi_bars}, who showed that the formation of bars in {\sc Auriga} is more common for galaxies that have higher stellar masses for a given halo mass, which appears to be the case for MW~910 (see section~\ref{auriga_comp_sec} in the Appendix, where we compare our Milky Way and M31 analogues to isolated galaxies in {\sc Auriga}).

Overall, while none of the simulated galaxies represent structurally close matches to the Milky Way across all properties, MW 1530 and MW 910 show some similarities with the Galaxy. Namely, MW~910 has a similar half-light radius \citep{lian_mw_size} and disc scale length to the Milky Way \citep{bland_hawthorn_gerhard}. MW 1530 is a closer match to Milky Way's disc mass \citep{disk_mass} and has a similar bar mass and length \citep{sourmani_bar, milky_way_bar_length}. The two galaxies also have higher contributions of the thin-to-thick disc, compared to MW 720 and MW 1540 and, as we will show in Section~\ref{sec_form_bim}, both show chemical bimodalities in their discs. In this work, we will therefore focus more closely on MW 1530 and MW 910.

\section{Galaxy assembly and the effect of mergers}
\label{sec:galaxy_assembly}

We now discuss the growth histories of our Milky Way analogues in dark matter and gas, their relation to the star formation histories of the galaxies and the relative contributions of mergers and smooth accretion to galaxy growth.

\subsection{Mass growth histories}

\begin{figure*}
    \includegraphics[width=2\columnwidth]{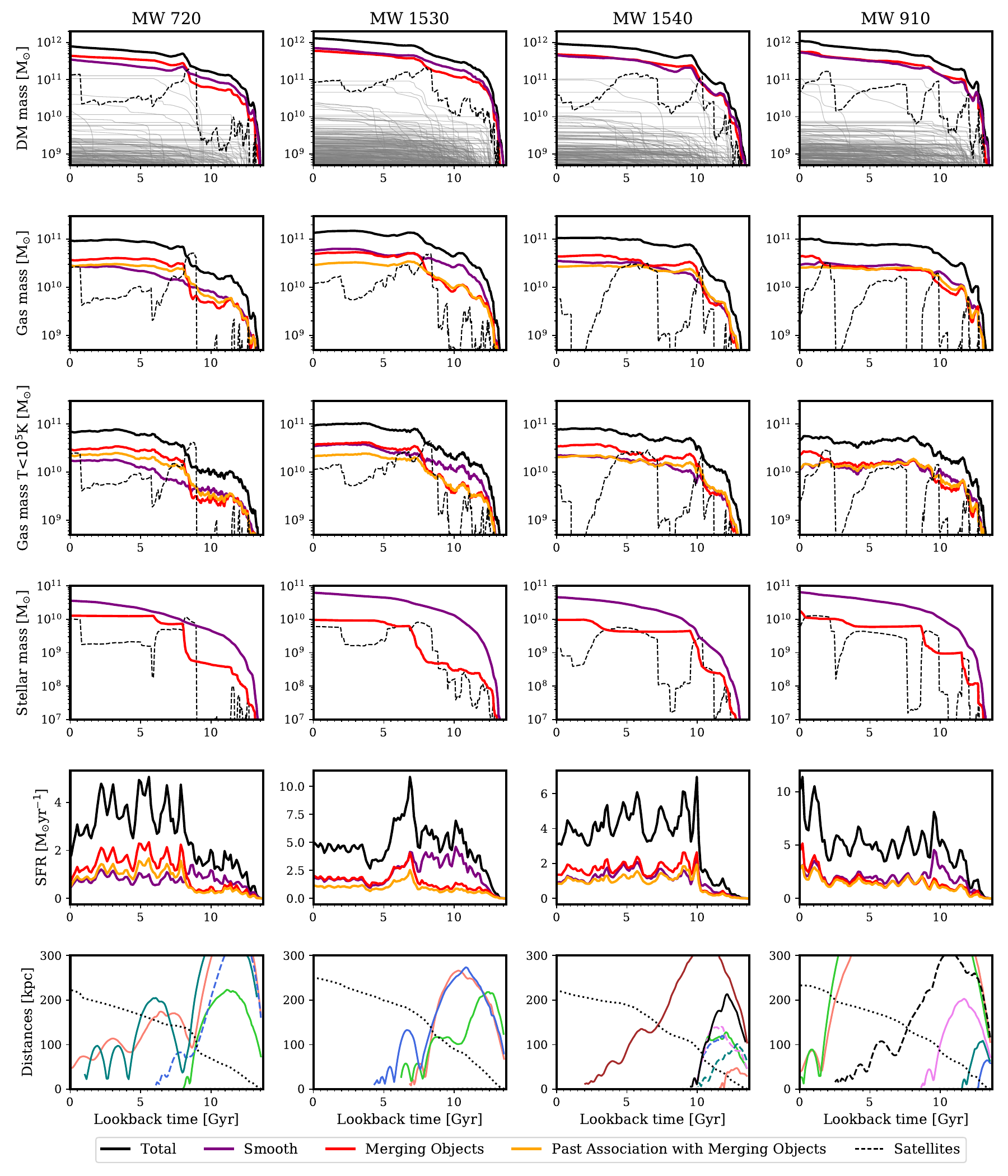}
    \caption{ The bound mass growth histories and star formation rates of each of our 4 Milky Way analogues. In the top three panels, the total mass growth (in black) is split in three contributions: smooth accretion (of dark matter and gas respectively, in purple), matter associated with mergers at the time of their maximum mass (red) and matter that had prior to infall some association with the merging objects (e.g. outflown gas, yellow). In the top row panels, the thin grey lines represent the mass deposited by each past merger with time. The black dashed lines show mass bound to all satellites that are within $R_{200}$ at the time. In case of multiple ongoing mergers, this mass can be comparable to or exceed that of the main galaxy. The second row shows the growth of the total mass in gas. The third row shows the cold gas mass, defined as the gas with temperature T < $10^5$~K. The fourth row shows the growth in stellar mass, plotting separately accreted stars in red. The fifth row shows the \textit{in-situ} star formation rate and relative contributions to it from the three sources of gas. The bottom panel shows the orbital evolution of past mergers and satellites with peak halo mass > 10~per~cent of the Milky Way analogue mass at that time. Solid lines show prograde interactions with respect to the disc, and dashed lines show retrograde interactions. The black dotted line shows the growth of $R_{200}$ (including mass in satellites). }
    \label{fig:mass-growth}
\end{figure*}

Figure~\ref{fig:mass-growth} shows the mass growth history and star formation rate of the four MW analogue galaxies depicted in Fig.~\ref{fig:present-day}. The top row shows the growth of the dark matter halo as a function of time, the second row the growth in the gas mass, the third row the content in cold gas (defined as that with $T < 10^5$~K)\footnote{We not that while this temperature cut separates well the gas heated in winds and AGN feedback and the starforming gas heated by the \citet{sh03} equation of state, at early times ($\approx$13~Gyr ago), this temperature is comparable to the virial temperature of the main halo, thus making the contributions of ``cold'' gas comparable to that of the ``hot'' gas.} and the fourth row the stellar mass. For the dark matter, we show the contribution of mergers alone (dark matter particles ever bound to mergers or satellites) with the red line, which is the sum of individual merger or satellite contributions (themselves indicated with thin grey lines), the smooth accretion\footnote{The smooth dark matter component here includes dark matter that was present at the time the halo appeared in the subhalo catalogue.} in purple and the total mass growth in black. For the gas, we subdivide the mass growth into three categories: contribution of mergers (which includes gas cells that belonged to the merging halo at the time of its maximum total mass), contribution of gas that was once bound to the merging halo but was not present at the time of maximum mass (which generally represents expelled gas), and the smoothly accreted gas that never belonged to a merging subhalo or a satellite. These sub-categories are shown in red, orange and purple, respectively. The fourth row shows the growth of the \textit{in-situ} stellar mass in purple and the accreted stellar mass in red. Across the top 4 panels the dashed black line shows mass in satellites located within $R_{200}$.

The fifth row shows the star formation rate inferred using the in-situ stars. For each newly-formed star we compute the contribution of each gas sub-component based on the origin of the gas tracers, where the contribution of the component $i$ corresponds to the mass of the star particle, $m_*$, times the tracer fraction $N_{\rm tracer, i}/ N_{\rm tracer, tot}$. Several of the Milky Way analogues appear actively star-forming at $z=0$, compared with a typical decline in the star formation rate seen in the {\sc Auriga} galaxies \citep{grand_auriga}. While we have verified that the $z=0$ star formation rate is within the scatter of the {\sc Auriga} galaxies (aside from MW 910), in Section~\ref{auriga_comp_sec} we show that our Milky Way analogues have slightly lower mass black holes compared to galaxies of similar stellar mass in {\sc Auriga}, which could account for their less efficient quenching.  

Finally, in the bottom panel we show the orbits of past mergers and satellites of each MW analogue with halo mass $>0.1M_{\rm MW}$.

\subsection{Dark matter}

Examining the growth in dark matter mass (top panels) for our sample of galaxies, we note the presence of big increases around the time that significant mergers take place. However, when comparing the mass associated with the mergers (red line) to the increase in mass of the main system (black line),  we find a significant difference between the two, made up by the smooth accretion of dark matter (purple line). The smoothly accreted component tends to dominate at early times. The difference can be as big as a factor of 2 - 2.5 (e.g. before the two mergers in MW 720) and in MW 1530, the smooth component dominates the mass contribution at all times. In MW 1540 and MW 910, the contributions of the two components appear fairly equal over the mass growth history. The smooth accretion is slightly enhanced during (or shortly after) mergers\footnote{We note that this result persists whether we consider bound mass according to {\sc SUBFIND-HBT} or mass within $R_{200}$.}. Some of this smoothly accreted component corresponds to dark matter haloes below our resolution limit, while some may represent dark matter that was never bound to a subhalo. These findings are in general agreement with previous studies \citep{wang_aquarius,edge}.

\subsection{Gas}

The growth of the gas component is in excellent correspondence with the dark matter growth histories, as shown in the second and third panels of Figure~\ref{fig:mass-growth}. Significant mergers cause a rapid increase in the gas mass, and are followed by a shallower increase in the smoothly accreted gas (purple line), which generally occurs with some delay after a merging halo entering $R_{200}$ (corresponding to a peak in the black dashed line). The delay is clear both in the total and cold bound gas, and typically 30~per~cent of the gas accreted following the merger is accreted in smooth form. This means that mergers also  {\it indirectly} lead to an increase in the gas reservoir available for star formation.   

Overall, the bound mass contributions of merger and smoothly accreted gas are often comparable (and reflect the relative contributions of the equivalent components in the dark matter). At late times, the gas mass contributed by mergers and satellites typically becomes dominant, while the contribution of smoothly accreted gas is more important at early times, when it also contributes substantially to star formation, as can be seen in the fifth row of the figure. This row also shows that around the time of the most massive mergers there is also a significant increase in the star formation rate. This effect is well-known, and has been reported in the literature before in simulations \citep{mihos_hernquist,cox,diMatteomerger,dimatteomerger2,moreno,li_torques} and observations \citep{sanders88, lambas, jogee, robaina,pearson, luo_starburst}.

The fluctuations in the cold gas content (third row) correlate well with the star formation rate (fifth row), with the relative contributions of smoothly accreted cold gas and that brought by or previously enriched by mergers reflecting their respective contributions to star formation. While the mass contribution of the different components to the star formation may differ, the separate bursts in the overall star formation rate are reflected in all three components, suggesting that they all contribute to the same episodes of star formation of the galaxy, although the chemical composition of the star-forming gas at any given time would be primarily influenced by the dominant contributor. We will discuss this further in Section~\ref{sec_form_bim}.

\subsection{Example of smooth accretion following an early merger}

\begin{figure*}
\includegraphics[width=2\columnwidth]{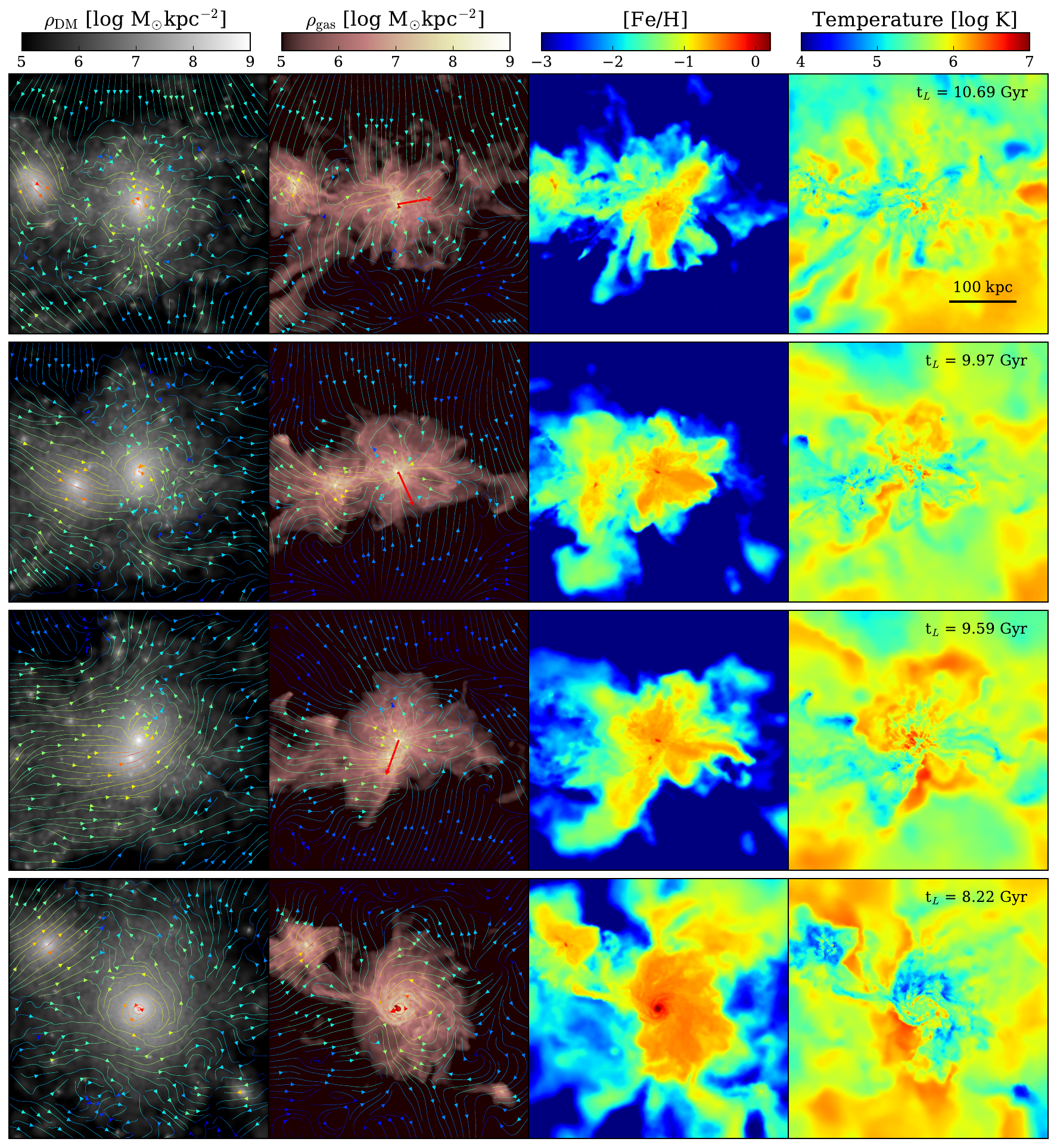}
\caption{The X-Z projection in simulation coordinates (approximately coinciding with a face-on projection with respect to the plane of the Local Sheet-like structure, see \citealt{hanneke_2026}), with the slice width $|Y| < 50$~kpc, of the dark matter (first column), gas (second column), gas metallicity (third column) and temperature (fourth column) of the Milky Way analogue 910 around the time of the GES-like merger, approximately 10 Gyr ago. The arrows show the dark matter and gas velocity field, centred on the stellar disc of the Milky Way analogue, with arrows pointing in the velocity direction and arrow colour indicating the logarithm of the mass flux (the product of density and the magnitude of velocity), with blue corresponding to low mass flux and red indicating high mass flux. The thick red arrows in the second column show the direction of the angular momentum vector of the gas disc. All properties are in physical units. }
\label{fig:snapshots}
\end{figure*}

Fig.~\ref{fig:snapshots} shows the distribution of dark matter and gas, gas metallicity, and temperature for galaxy MW 910 around the time of the merger, which we may be considered an analogue of Gaia-Enceladus, with halo mass of $10^{11} M_{\odot}$ and DM mass ratio to the Milky Way analogue of 2.5~:~1 (gas ratio 2~:~1), entering the virial radius 10~Gyr ago. To follow the flow of dark matter and gas, we also display the velocity field, computed in physical units in the $100^3$~kpc$^3$ box and weighted by mass. The velocities are centred on the centre-of-mass velocity of the stellar disc, computed using the stellar particles in the inner 10~kpc.

In the top panel of Figure~\ref{fig:snapshots} we show the X-Z projection in the simulations coordinates, corresponding to an approximate face-on view\footnote{\citet{hanneke_2026} have found that the Y-coordinate in these simulations coincides within~10 degrees with the normal of the plane fitted to the Local Sheet-like structure.} with respect to the plane of the Local Sheet-like structure (where the Milky Way and M31 analogues lie in the Local Group realizations of \citealt{Wempe2024,wempe_sheet}). Note the inflow of both gas and dark matter on the left side of the plot, that is associated with the incoming GES-like merger. Mass is also accreted onto the Milky Way analogue from other directions within the Local Sheet structure, indicated by the velocity field above and below. As the GES analogue enters within 100~kpc of MW 910 (third row, 9.59~Gyr ago), we see increased inflow from the original direction of the merger both in the gas and the dark matter, as more matter is drawn into the deepened potential well of the MW 910. The inflowing gas is more metal-poor and colder than the halo gas. Finally, as the merger is complete and the next important merger ($M_{\rm halo} = 8\times10^{10}M_{\odot}$, $M_{\rm halo}/M_{\rm MW} = 0.25$) is approaching on the upper left of the fourth panel (8.22~Gyr ago), the gas is preferentially outflowing with several bubbles blown by the winds and AGN feedback. The Milky Way's disc is seen face-on in this orientation and contains a significant fraction of cold gas. Interestingly, in this projection, we see residual ordered motion in the dark matter following the merger that is roughly in the plane of the disc, as indicated by the direction of the angular momentum vector (black arrow). The alignments between the angular momentum vectors of galaxies and their dark matter haloes in our suite of simulations are discussed in a companion paper \citep{hanneke_2026}.

\section{Formation of chemical bimodality}
\label{sec_form_bim}

\begin{figure*}
\centering
\includegraphics[width = 2\columnwidth]{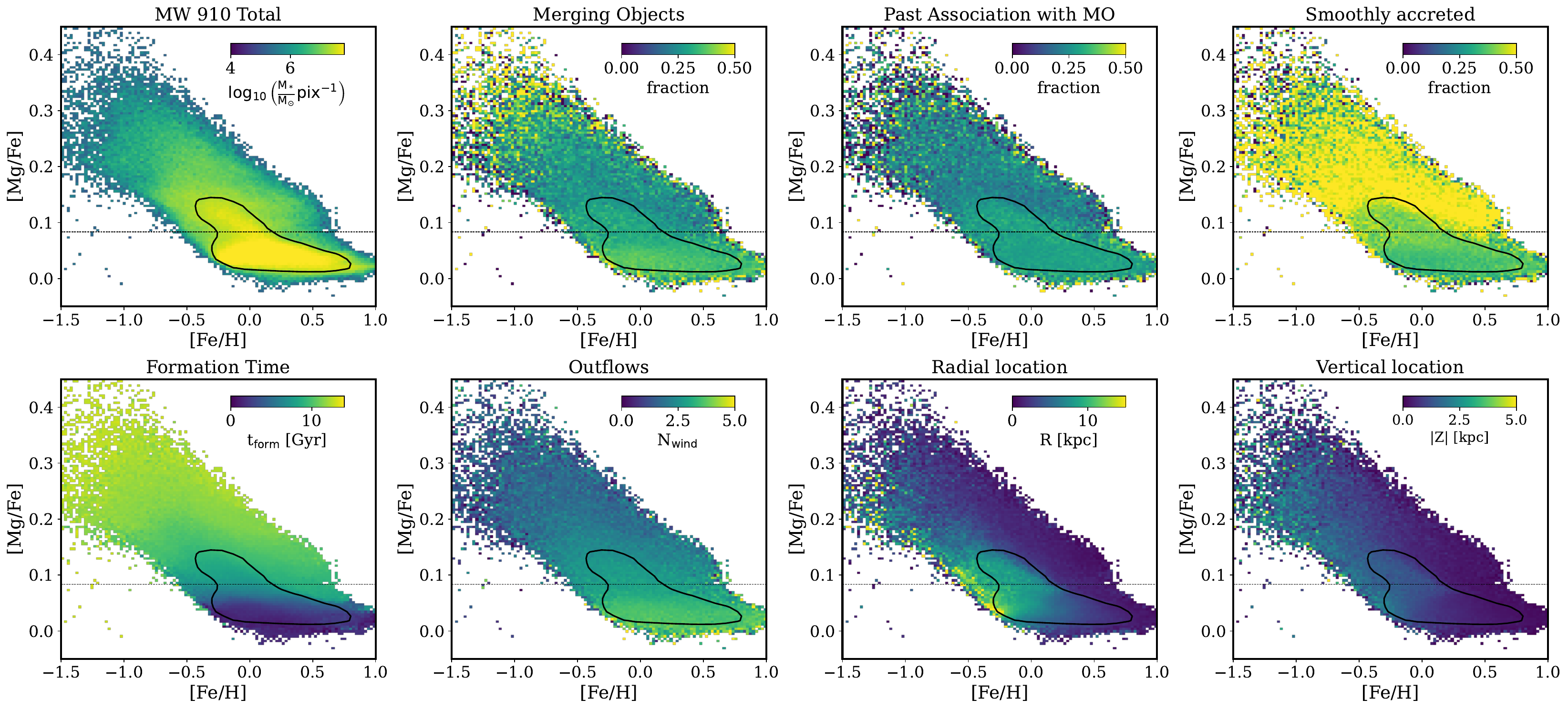}
\caption{The [Mg/Fe] - [Fe/H] distribution for all stars formed \textit{in-situ} in MW 910. The top left panel shows the total distribution of stellar mass on the diagram, while the other panels in the top row show the fraction of mass in each [Mg/Fe]-[Fe/H] bin from gas contributed by merging objects (counting all tracer particles bound to the objects at the time of their maximum gas mass), the gas ejected from merging objects prior to infall onto the Milky Way analogue, and the smoothly accreted gas. The bottom row shows, from left to right: the average formation time of stars in each bin, the average number of times each gas tracer particle was launched in a wind, the radial location of the stars at the present day and the vertical location of the stars. The stellar particles were selected to have a maximum radial location of 20~kpc and a maximum vertical height of 5~kpc. The black contour outlines 68~per~cent of the probability distribution and the low and high-$\alpha$ populations are separated within this contour with the black dashed line.}
\label{smooth_gas_props}
\end{figure*}

In this section, we explore how the various channels of gas accretion (via merging satellites or smoothly accreted gas) have contributed to the chemo-dynamical properties of our Milky Way analogues. We then study the chemical composition of the discs of our simulated galaxies, the presence of bimodalities within them, and their origin. 

\subsection{Origin of gas forming the chemical sequences}

Following the discussion on the enhanced accretion of smooth gas following a merger in MW 910, in Fig.~\ref{smooth_gas_props} we show the chemical composition of the \textit{in-situ} stars in this galaxy. Like in the {\sc Auriga} simulations, our MW galaxies' mean metallicities are systematically higher than observed, although within the scatter of the mass-metallicity relations \citep{grand_auriga_new,grand_auriga_release}. Furthermore, to compensate for underproduction in the \citet{portinari} yields we have used, we have rescaled the Magnesium abundance in our simulations by a factor of 2.5 following \citet{van_de_voort}. 

In the top left panel of Fig.~\ref{smooth_gas_props}, we see a broad distribution in metallicity and [Mg/Fe] abundance with two prominent sub-components (containing 68\% of the stars, black contour) resembling the [$\alpha$/Fe] bimodality observed in the Milky Way. The top three panels to the right show the fraction of stellar mass in each bin of [Mg/Fe]-[Fe/H] made up by tracers from three gas acquisition channels: merging objects, gas previously associated with merging objects, and smoothly accreted gas. Smooth accretion is the largest contributor to the formation of the most $\alpha$-enhanced stars in the simulated galaxy, some of which are relatively metal-rich. However, it is clear that the two sequences outlined in the leftmost panel are made up of the nearly equal combination of all three channels. Specifically, in the high-$\alpha$ sequence we find fractional contributions of mergers, merger-associated and smoothly accreted gas of 0.30, 0.29 and 0.41, respectively, while in the low-$\alpha$ sequence this is 0.35, 0.29 and 0.36, respectively.

In the bottom row of Fig.~\ref{smooth_gas_props}, we show, from left to right, the average formation time of the stars in each bin, the average number of times the tracers contributing to star formation were launched in a wind, as well as the radial and the vertical location of the stars in the disc at the present day. The two rightmost panels show that the most metal-rich stars are centrally concentrated, while the most metal-poor stars are generally seen at large distances and heights above the disc, suggesting that these stars formed in the outskirts of the galaxy. 

Comparison of the top rightmost and the bottom leftmost panels reveals that the stars made up primarily of smoothly accreted gas have formed early on in the history of the galaxy. These earliest forming stellar populations (> 11~Gyr old) have a non-negligible contribution from cold flows (see  Fig.~\ref{hotcold} in the Appendix). In contrast, the high- and low-$\alpha$ sequences in this MW analogue seem to have formed later in the history of the galaxy. Most of the stars in the high-$\alpha$ sequence formed between 9.6 and 7.1 Gyr ago, i.e. after the merger with the GES analogue. The low-$\alpha$ population instead formed relatively recently, with most of the stars being younger than $\sim 5$~Gyr. Both populations have formed from gas that was launched in winds multiple times. 

The high-$\alpha$ sequence had an important contribution (close to 40\%) from gas from the GES-like object, while the low-$\alpha$ sequence has comparable contributions from GES and the Sagittarius-like object (roughly 20\% each, see Fig.~\ref{sag_ges} of the Appendix \ref{app:GES-Sgr-gas}. Note that this figure suggests  
a weak imprint of the metallicity of the merged system is left on the sequence that forms after). 
The gas forming the high-$\alpha$ sequence typically spends $2.63^{+1.61}_{-1.15}$~Gyr in the halo before forming stars, while the gas forming the low-$\alpha$ population spends $6.43^{+2.96}_{-2.92}$~Gyr in the halo, indicative of hot-mode gas accretion \citep[as also found in][]{grand2019gas}. 

We also see that stars on the high-$\alpha$ sequence are more radially extended than the low-$\alpha$ ($\langle R_{\rm \alpha,high}\rangle = 7.27\pm0.03$~kpc, compared to $\langle R_{\rm \alpha,low}\rangle = 5.99\pm0.01$~kpc), with higher typical heights above the midplane ($\langle |z|_{\rm \alpha,high}\rangle = 1.24\pm0.01$~kpc, compared to $\langle |z|_{\rm \alpha,low}\rangle = 0.86\pm0.01$~kpc), however sub-solar metallicity stars in the low-$\alpha$ population are typically more spatially extended.

\begin{figure*}
\begin{multicols}{2}
\includegraphics[width = \columnwidth]{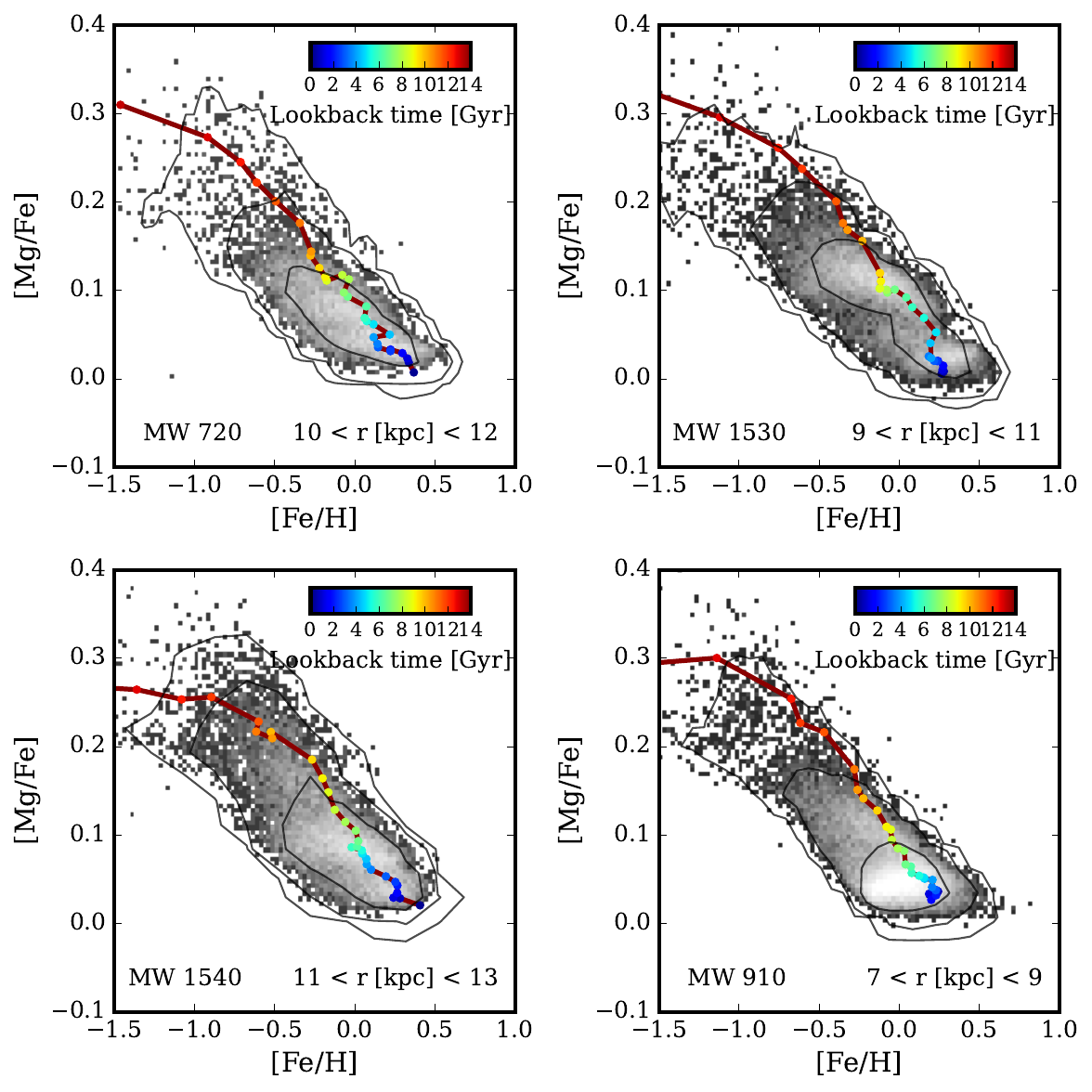} \\
\includegraphics[width = \columnwidth]{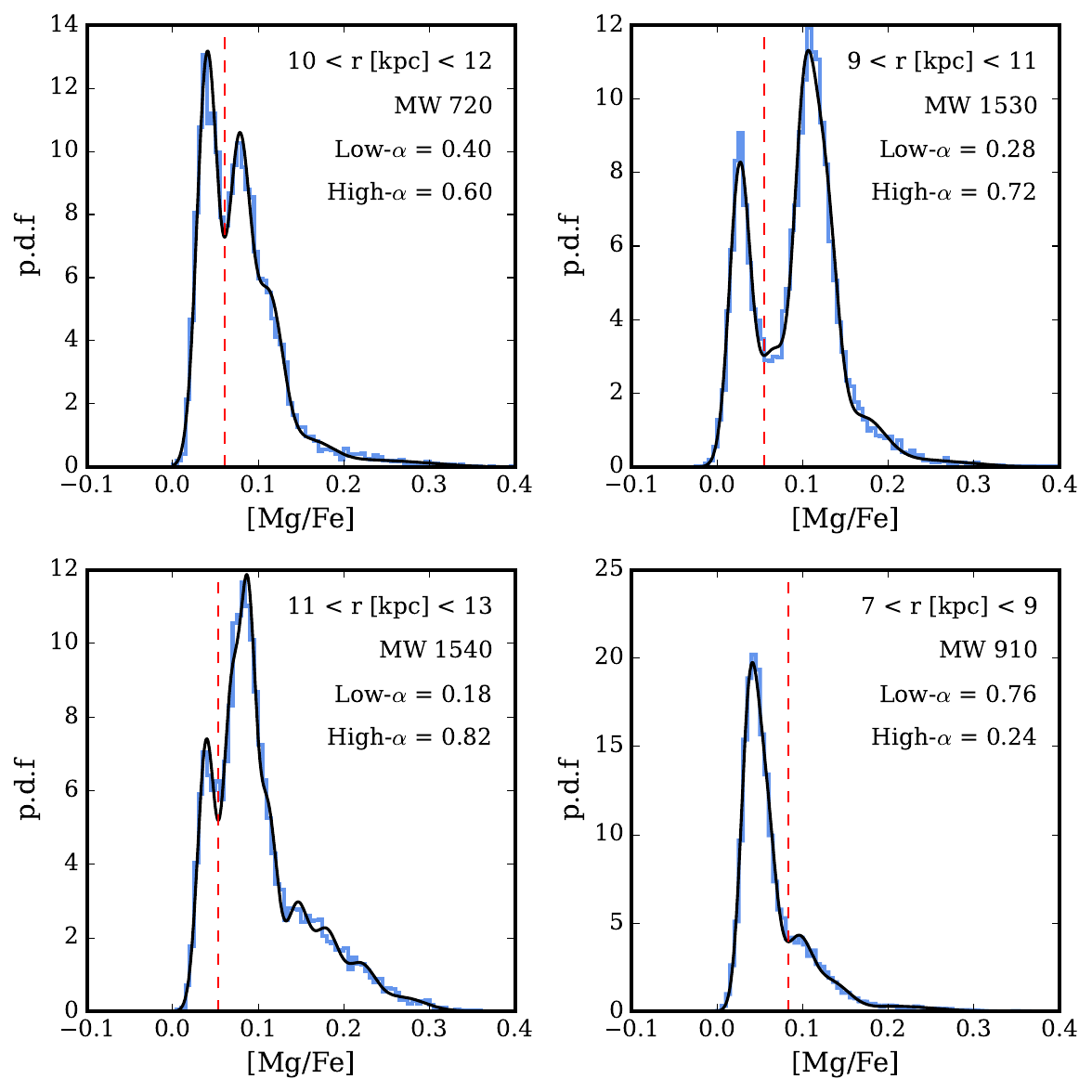}
\end{multicols}
\caption{\textit{Left:} [Mg/Fe] -- [Fe/H] distribution for stars located within 1~kpc of the rescaled Solar radius in the plane of the disc, with vertical heights |$z$|<5~kpc, of our 4 Milky Way analogues. The contours show the 68$^{\rm th}$, 95$^{\rm th}$ and 99$^{\rm th}$ percentiles of the density distribution. The dark red line joining the coloured dots shows the median locations on the diagram of the stars formed in each age bin, with lookback time of formation indicated in the colour bar. \textit{Right:} The [Mg/Fe] distribution of stars located within 1~kpc of the rescaled Solar radius in the disc plane, with |$z$|<5~kpc of the 4 Milky Way analogues. The black lines show the Gaussian mixture model fits to the distributions, from which we define a cut between the $\alpha$-rich and the $\alpha$-poor populations in this radial bin (red dashed line). The labels show the fraction of stars belonging to each population. }
\label{fig_8_kpc}
\end{figure*}

In summary, we have seen that 
gas acquired via mergers and via smooth accretion contribute in similar amounts to the formation of stars in the disc of MW~910. We have also identified $\alpha$-sequences but their origin does not appear to be exclusively related to either of these two gas acquisition modes.  In the next Section, we will examine in more detail the presence of distinct $\alpha$ sequences across the four MW analogues and establish their physical origin.

\subsection{$\alpha$-Fe sequences in the Solar neighbourhood}

On the left of Fig.~\ref{fig_8_kpc}, we show [Mg/Fe]-[Fe/H] distributions of \textit{in-situ} stars in the "Solar neighbourhood" of the 4 Milky Way analogues. The rescaled Solar radius is obtained by deriving the $r$-band luminosity half-light radius for stellar particles within 17~kpc of the galactic centre and multiplying by a factor of 1.4, equivalent to the position of the Sun with respect to the Milky Way's half-light radius as derived in \citet{lian_mw_size}. The derived half-light radii are listed in Table~\ref{tab:main-props2}. The red line with coloured dots in Fig.~\ref{fig_8_kpc} marks the median abundances at which stars form at each lookback time. These tracks show the expected increase in [Fe/H] and decrease in [Mg/Fe] over time, as well time periods over which the metallicity of the galaxy is temporarily reset. As we will see below, these metallicity resets are related to the timing of significant accretion episodes. On the right, we display the distributions of [Mg/Fe] for each galaxy, which we fit with a combination of up to 10 Gaussian components \citep{gmm}, primarily with the aim to define the cut between the low-$\alpha$ component and the higher $\alpha$ components, which is taken as the location of the first minimum in the Gaussian mixture model. The number of Gaussian components is determined using the Bayesian Information Criterion \citep{bayesian}. From the peaks in the distribution, it can be seen that at least 2 distinct $\alpha$-populations exist in each of our analogues at this radial distance. The prominence of the low-$\alpha$ sequence varies from simulation to simulation, being the most dominant for MW~910 and least for MW~1540.  

MW 720 displays the presence of at least 3 distinct $\alpha$-populations. This galaxy has the highest contribution of merger gas to the star formation history of all four MW analogues, and this may explain the more complex changes in the chemical abundances of newly formed stellar populations ("wiggly" evolution). However, the quick succession in which these changes occur results in relatively small gaps between successive peaks in the [Mg/Fe] distribution, while the dilution does not appear to be sufficient to significantly reset [Fe/H] in the newly formed $\alpha$-sequences. This result is consistent with the work of \citet{bustamante_mergers}, who, studying {\sc Auriga} simulations, do not find dilution greater than $\approx$0.15~dex, except for mergers with mass ratio close to unity, or if the merging pair is affected by a third galaxy.

A clear bimodality is visible in the case of MW~1530, with the track outlining the chemistry of newly-formed stellar populations showing some ``loops'' (see also Fig.~\ref{fig:chemistry}). The high-$\alpha$ sequence is more dominant than the low-$\alpha$ sequence, contrary to what is observed in the MW. In comparison, MW~1540 shows a similar overlap in metallicity in the two $\alpha$-sequences and a less pronounced gap in $\alpha$. The high-$\alpha$ sequence in this galaxy is substantially more massive than in the other MW analogues. In Section~\ref{m31_720_sec} of the Appendix, we also discuss M31 720, where a greater burst in star formation and no subsequent satellite or merger interaction leads to 
two $\alpha$-sequences of similar extent as the Milky Way's, and which are separated by a strong [Mg/Fe] gap as also observed.

Finally, MW 910 shows a bimodal distribution of [Mg/Fe] (as in Fig.~\ref{smooth_gas_props}), although the gap between the two $\alpha$ populations is less significant than in MW 1530. On the other hand, there is a more significant overlap in [Fe/H] of the low and high-$\alpha$ sequences.

\subsection{Star formation history across the disc}

\begin{figure*}
\includegraphics[width=\columnwidth]{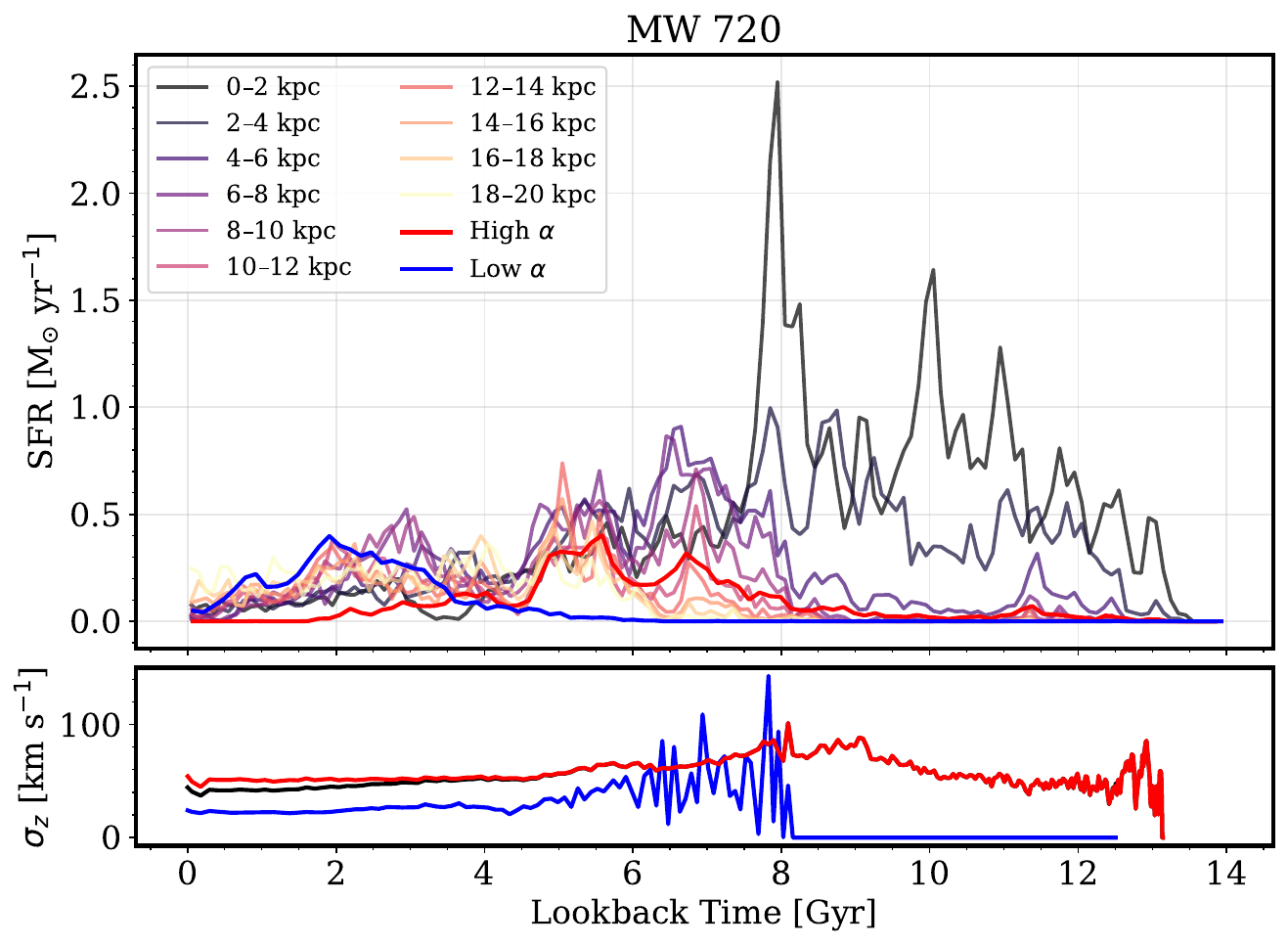}
\includegraphics[width=\columnwidth]{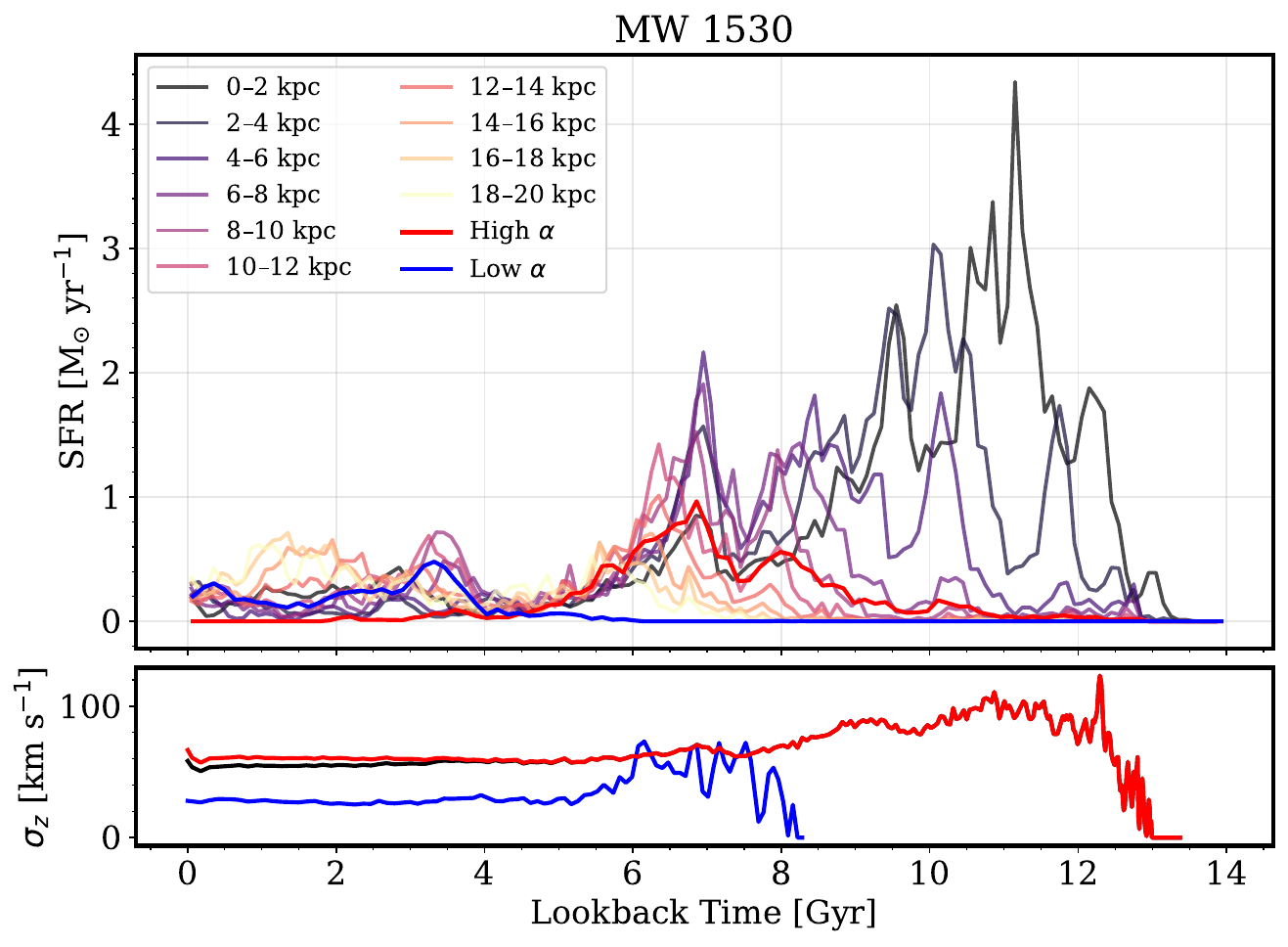} \\
\includegraphics[width=\columnwidth]{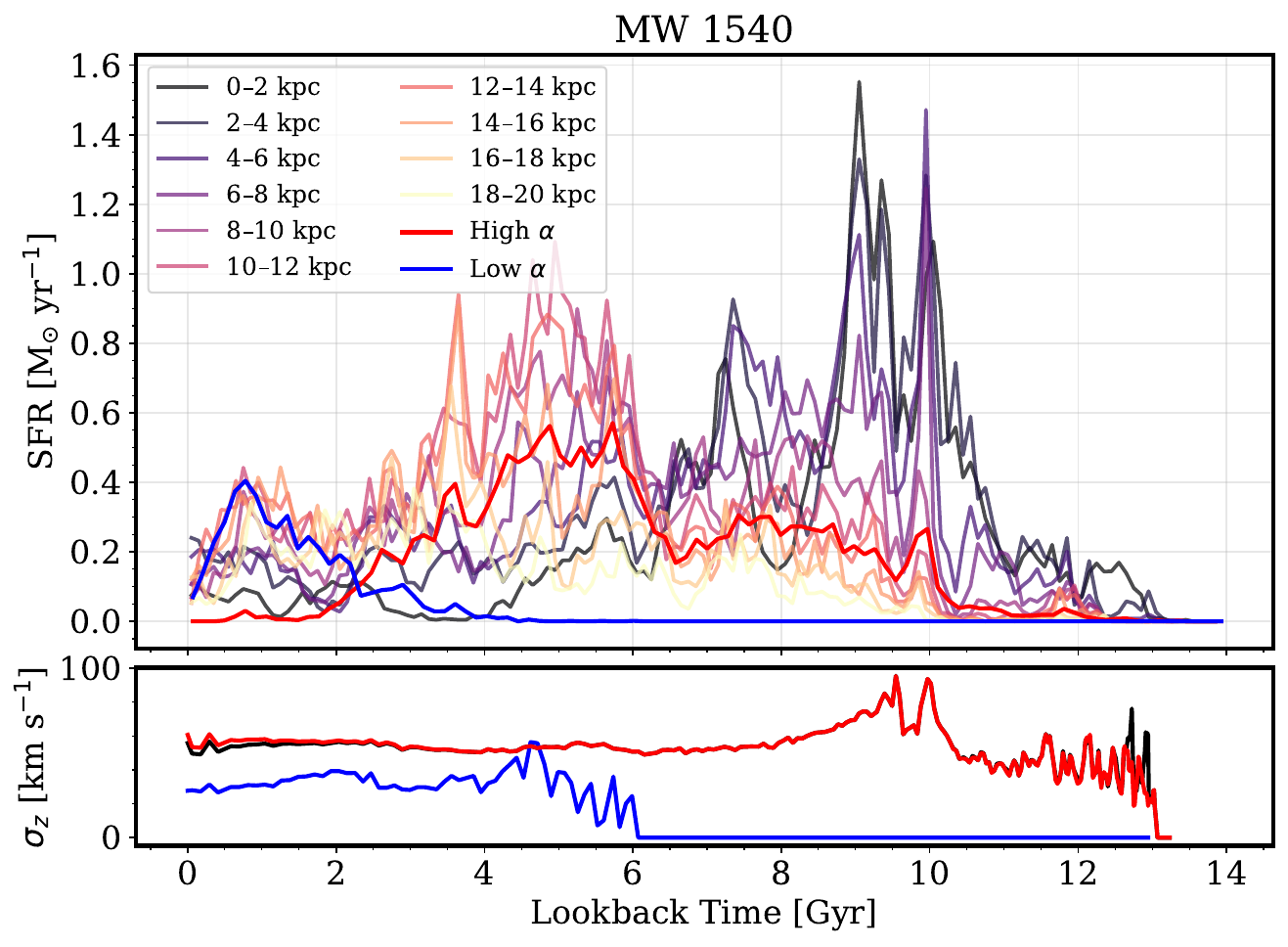}
\includegraphics[width=\columnwidth]{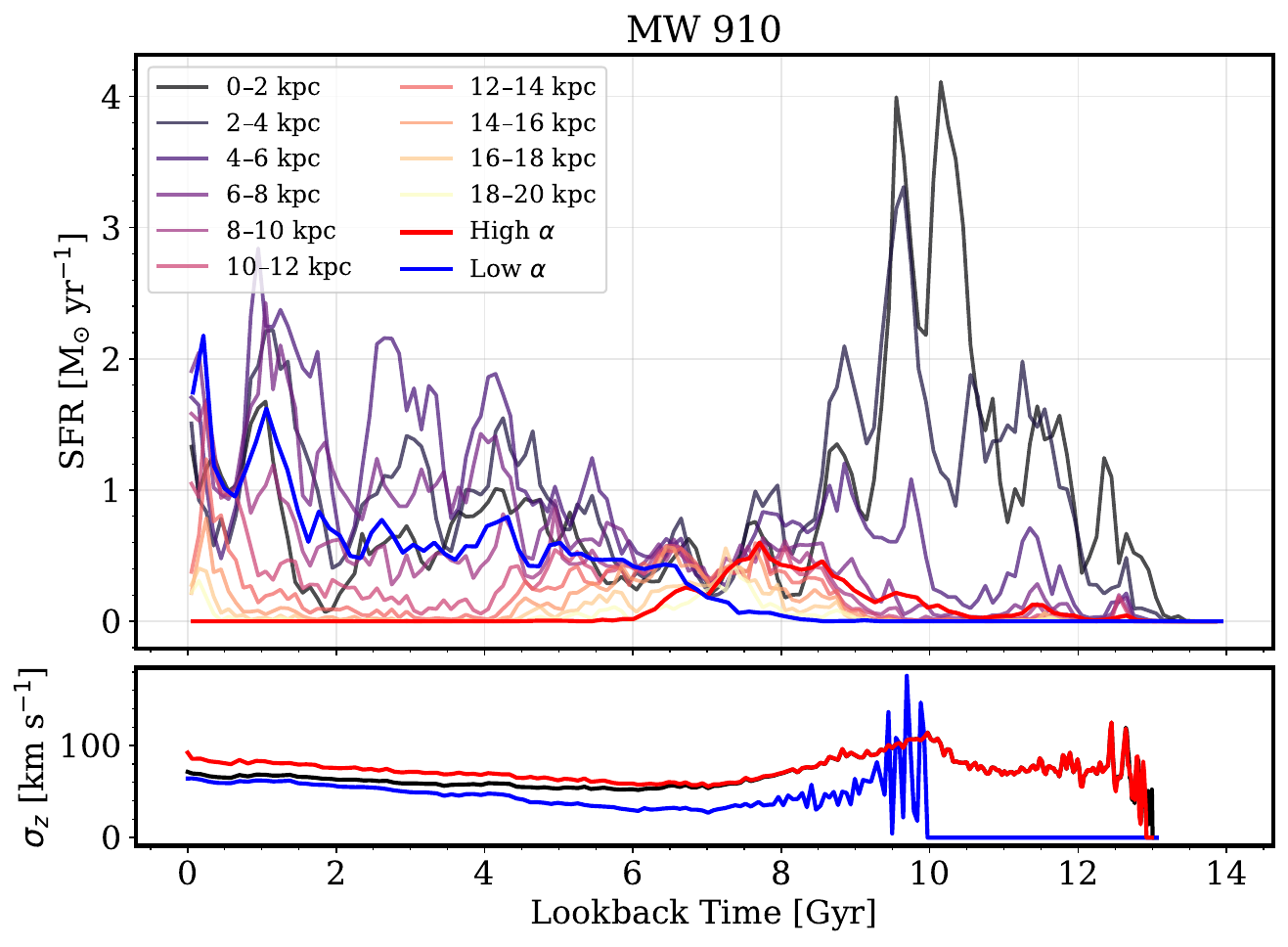}
\caption{\textit{Top panels:} The star formation rate history in 2~kpc radial bins, with the vertical height $|z|<5$~kpc across the discs of the Milky Way analogues. The radial bins can be identified by their colour in the legend. The red and blue lines show the star formation rate history for the $\alpha$-rich and the $\alpha$-poor populations, respectively, for stellar particles which are within 1~kpc of the rescaled Solar radius at present day, selected with a rigid [Mg/Fe] cut as in Fig.~\ref{fig_8_kpc}. \textit{Bottom panels:} the vertical velocity dispersion as a function of time for all stars which are at the present day within 1~kpc of the rescaled Solar neighbourhood radius as in Fig.~\ref{fig_8_kpc} (black) and the $\alpha$-rich (red) and $\alpha$-poor (blue) populations.}
\label{fig:sfr_disk}
\end{figure*}

We now examine in more detail the star formation history across the discs of the Milky Way analogues and the kinematics of the $\alpha$-rich and $\alpha$-poor sequences. Previous studies (e.g. \citealt{orkney, barry_fire}) have shown that the formation of a low-$\alpha$ sequence occurs as a result of the decline in star formation, allowing ongoing Type~Ia SNe to enrich the ISM, lowering [Mg/Fe] for new stellar populations when star formation resumes. Fig.~\ref{fig:sfr_disk} shows the star formation history in bins of 2~kpc across the galactic disc, while red and blue lines show the star formation histories for the $\alpha$-rich and $\alpha$-poor populations, respectively, within 1~kpc of the rescaled Solar neighbourhood radius at the present day. The two populations are defined by a rigid [Mg/Fe] cut, as shown in the right-hand panel of Fig.~\ref{fig_8_kpc}.  

Despite the star formation histories in Fig.~\ref{fig:mass-growth} showing no significant gaps for all galaxies aside from MW 1530, in Fig.~\ref{fig:sfr_disk}, we see that the start of the formation of the $\alpha$-poor sequence often coincides with a relative decline in star formation for many radii across the galaxy. The decline typically follows a burst of star formation during which a high-$\alpha$ sequence forms. This is particularly prominent in MW 1530 and the lowest-$\alpha$ populations in MW 720 and MW 1540, which are reminiscent of star formation histories of $\alpha$-bimodal {\sc Auriga} galaxies in \citet{orkney}. It is less clear for MW 910, where the star formation rate only drops slightly and less globally in the transition between the $\alpha$-rich and the $\alpha$-poor sequences. Instead, the star formation rate seems to drop only in the inner parts of the galaxy (below 8~kpc) in MW~910 7~Gyr ago, which is likely to result in some radial dependence of the observed bimodality \citep[as also seen in][]{grand_discs}. MW 910 has an additional reduction of star formation in the outer parts of the galaxy (>8~kpc) between 5 and 2~Gyrs ago.

In the subpanels of Fig.~\ref{fig:sfr_disk} we show the vertical velocity dispersion in the stars of the low and high-$\alpha$ populations in the four MW analogues. It is clear across the four simulations that the low-$\alpha$ population is kinematically colder than the high-$\alpha$ population, suggesting the low-$\alpha$ population is more likely to make up the kinematic thin disc. In MW 720 and MW 1530, the low-$\alpha$ population and thin disc form after the star formation becomes relatively quiescent, as also seen in the analysis of {\sc FIRE 2} simulations in \citet{yu_thin}. In MW~1540 and MW~910, the velocity dispersion rises in both components and the difference between them is somewhat narrowed, suggesting kinematic heating, which is more significant for the low-$\alpha$ disc. As can be seen in Fig.~\ref{fig:mass-growth}, this is the result of an extended interaction and a final merger with a Sagittarius-like satellite for MW 910.

\begin{figure*}

    \centering
    \includegraphics[width=\columnwidth]{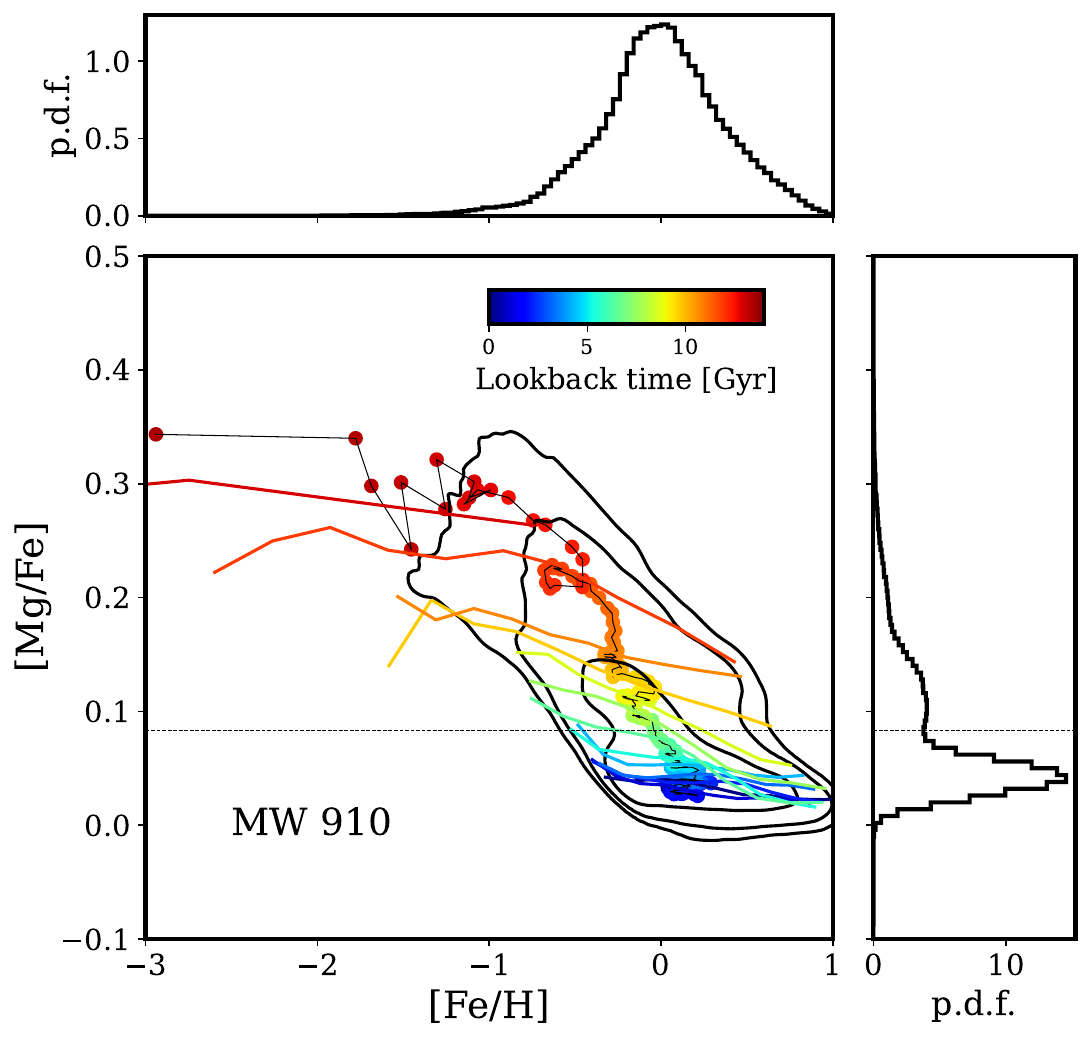}
    \includegraphics[width=\columnwidth]{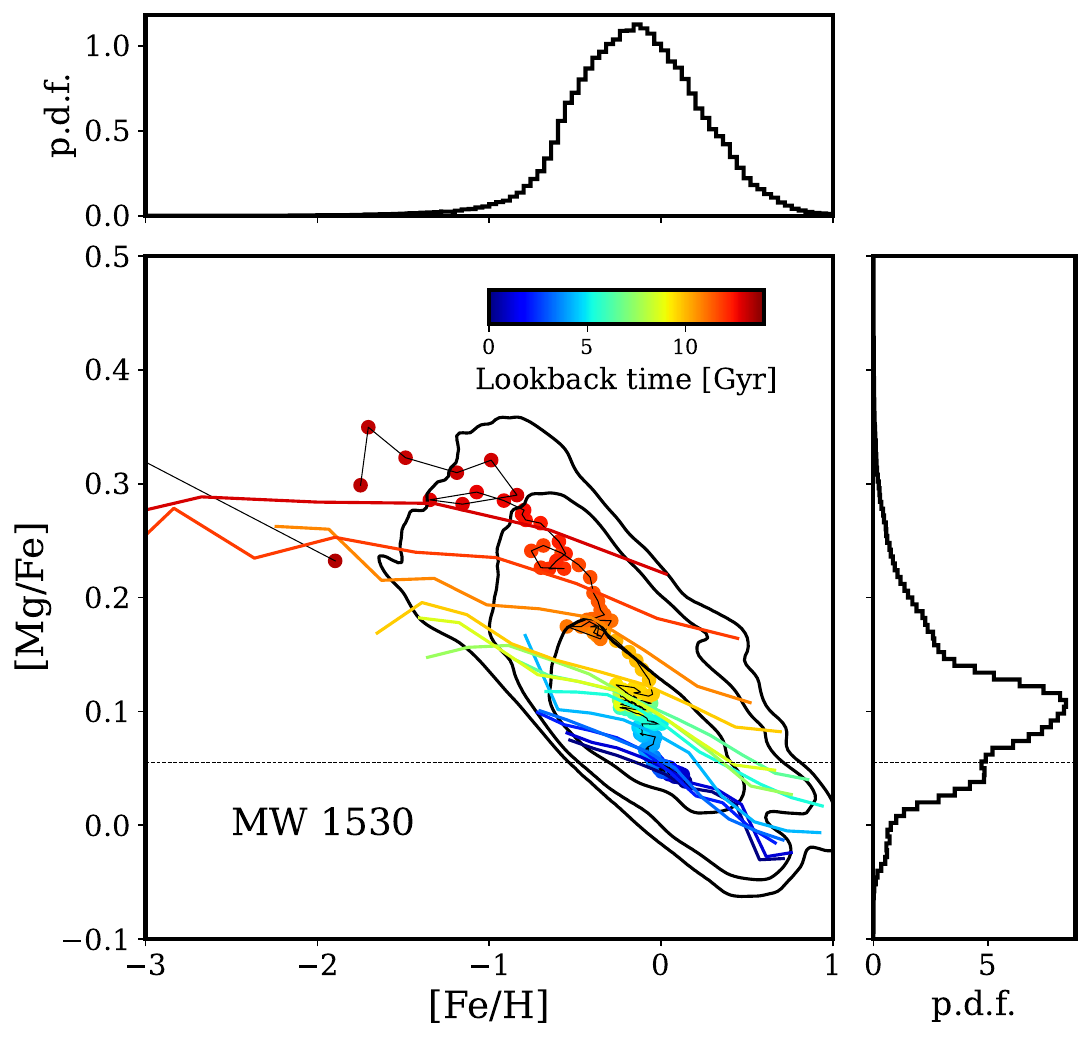}

    \caption{The time evolution of the [Mg/Fe]~-~[Fe/H] diagram for all \textit{in-situ} stars of MW 910 (left) and MW 1530 (right). The black contours show the 68$^{\rm th}$, 95$^{\rm th}$, and 99$^{\rm th}$ percentiles of the distribution of the stars. The thin black line with coloured dots follows the median [Mg/Fe] and [Fe/H] of stars in each age bin, as indicated by the colour bar. The mean metallicity of newly-formed stars executes several ``loops'', which coincide with past merger events. The coloured lines show mean [Mg/Fe] for each bin in [Fe/H] of stars forming at a given lookback time. The panels at the top and the right show the marginal distribution of [Fe/H] and [Mg/Fe], respectively. The black dotted lines show the cut used to separate the low and high--$\alpha$ populations in Fig.~\ref{fig_8_kpc}.}

    \label{fig:chemistry}
\end{figure*}

\subsection{Different types of [$\alpha$/Fe] bimodalities and their origin}

\begin{figure*}
\begin{multicols}{2}
\includegraphics[width = \columnwidth]{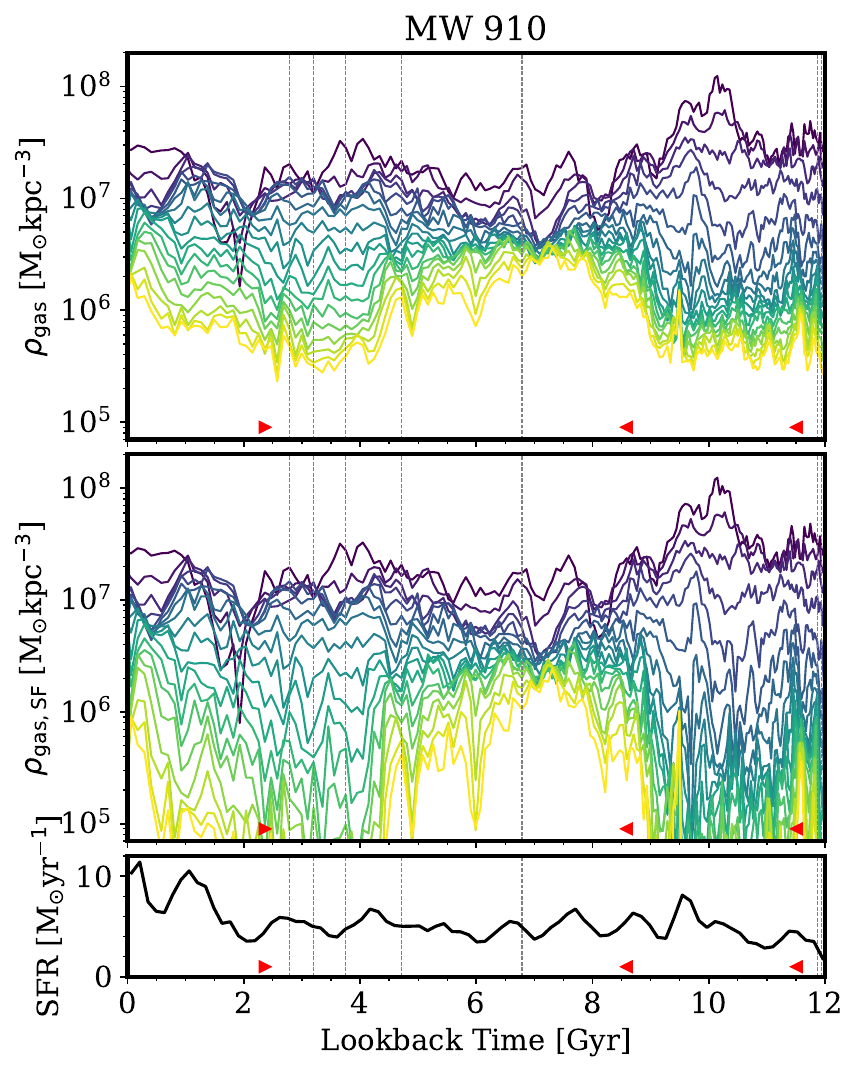}\\
\includegraphics[width = \columnwidth]{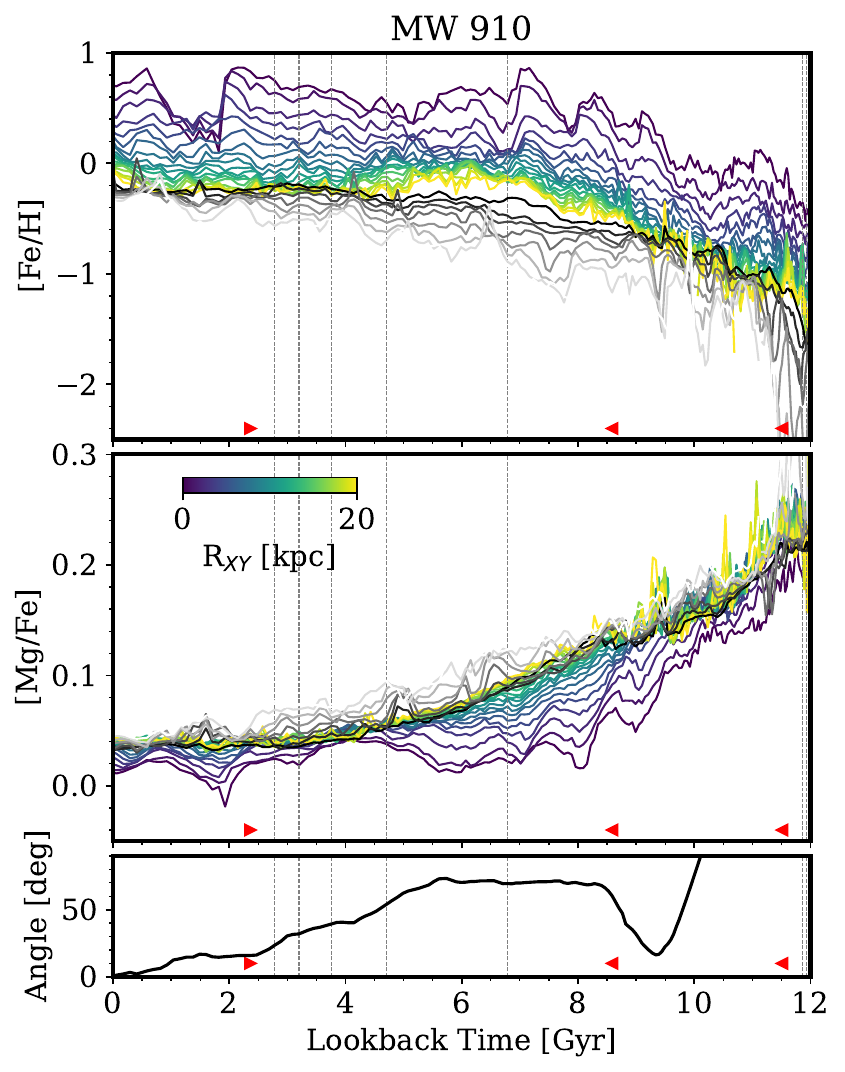}
\end{multicols}
\caption{\textit{Left:} The top panel shows the gas density across the disc of MW 910, with each line corresponding to a radial bin from 0 to 20~kpc (as identified with the colour bar) and $|z|<5$~kpc. The middle panel shows the density of star-forming gas across the disc. The dashed vertical grey lines show the pericentres of the Sagittarius-like dwarf. The red triangles show past merger events, with left-pointing triangles indicating prograde mergers and right-pointing triangles indicating retrograde mergers. The bottom panel shows the star formation rate for all \textit{in-situ} stars in the galaxy. \textit{Right:}[Fe/H] (top) and [Mg/Fe] (middle) of star-forming gas in each radial bin across the disc. The grey lines show [Fe/H] and [Mg/Fe] of the CGM in 8 bins from 20 to 100~kpc (darker to lighter grey lines), computed in edge-on radial bins with respect to the disc, with vertical height of $|z|<5$~kpc. The bottom panel shows the difference in angle between the angular momentum vector at a given lookback time and the angular momentum vector at the present day. After the arrival of Sagittarius the disc in MW 910 tilts (see \citealt{dillamore_tilting, dodge_tilting}) and the entire star-forming disc has almost a uniform composition in [Mg/Fe], while exhibiting a radial gradient in [Fe/H]. }
\label{910_disk_fig}
\end{figure*}

In the following, we look more closely at MW 910 and MW 1530, both of which show [$\alpha$/Fe] bimodality, while exhibiting different star formation and merger histories. Fig.~\ref{fig:chemistry} shows the [Mg/Fe]--[Fe/H] distribution of all \textit{in-situ} stars in the two MW analogues and the marginal distributions of [Mg/Fe] and [Fe/H]. The dotted line follows the median chemical properties of successively forming stellar populations, while the coloured lines show mean [Mg/Fe] for each bin in [Fe/H].

For both galaxies, we see more prominent resetting of the metallicity, resulting in clearer ``loops'' in the evolutionary tracks than in Fig.~\ref{fig_8_kpc}. These are particularly clear at early times (> 9 Gyr) and correspond to mergers in the history of the MW analogues. We also see that for both galaxies, stars form across a wide range of metallicities at each epoch. In MW 1530, there is also a strong gradient in [Mg/Fe] with respect to [Fe/H] at a given age, with more metal-rich stars being more $\alpha$-poor and vice versa. 
 
The distribution in MW 910 displays a clear bimodality in [Mg/Fe] and a more prominent low-$\alpha$ peak, but unlike MW 1530, this galaxy shows little to no gradient in [Mg/Fe] for stars formed between 5~Gyr ago and the present day (when the low-$\alpha$ sequence forms in Fig.~\ref{fig:sfr_disk}), suggesting that the Mg-abundance changes at the same rate as the Fe-abundance at these times. Instead, at earlier times, there is a clear gradient in [Mg/Fe] across the disc, similar to what is seen for MW 1530. The clear difference in the [Mg/Fe] slopes of the $\alpha$-poor sequences in the two analogues reflects their distinct origin, as we will discuss in the following\footnote{In Sec.~\ref{resolution_910_sec} we also show the bimodality in the lower-resolution realization of MW 910, which displays broadly similar properties to the high-resolution version.}. 

\subsubsection{Origin and extent of the bimodality in MW 910}

We investigate the mechanism through which the bimodality emerges further in Fig.~\ref{910_disk_fig}. On the left, we show the density of gas across the disc, out to 20~kpc, and the density of star-forming gas at the same radii. The dotted grey lines mark the pericentres of the Sagittarius analogue in this simulation and red triangles show merger times (defined as the snapshot after the merging galaxy was last identified with the subhalo finder). As the Sagittarius analogue approaches its second pericentre ($\sim$ 40~kpc from the galaxy), the gas density drops in the outer disc, with a stronger effect in the star-forming gas, indicating that the interaction tends to puff up the gas disc, also explaining the heating of the stellar populations seen in Fig.~\ref{fig:sfr_disk}. The bottom right subpanel shows the orientation of the disc compared to that of today. On the second approach of the Sagittarius analogue, the disc also begins to tilt, moving by nearly 70 degrees to its present day position. In the innermost regions of the galaxy, the gas density increases during the first pericentric passage of the Sagittarius analogue, leading to a burst in star formation. The density also increases as the Sagittarius approaches its second apocentre, which leads to another peak in star formation $4.2$~Gyr ago (left bottom subpanel).

The orbit of our Sagittarius analogue in MW 910 is retrograde with respect to the disc's rotation (with an initial orbital inclination of $\approx 60$~degrees, reducing gradually to zero from the first apocentre after infall), thus reducing the angular momentum in the outer disc of the Milky Way analogue. This is reminiscent of the formation of chemical bimodality seen in the previous studies by  \citet{johnson_retro} and \citet{amarante}, who investigated the impact of a pro- and retrograde GES-like merger on the formation of distinct $\alpha$-Fe sequences. In our case, the GES-analogue has a gas mass ratio of 2~:~1 (dark matter ratio 2.5~:~1), is prograde, and leads to the formation of the high-$\alpha$ sequence over a relatively extended period of time. Therefore, in the case of MW~910, it is not the GES-analogue but the retrograde Sagittarius analogue, with gas mass ratio of 5~:~1 (dark matter ratio 5~:~1), that reduces the angular momentum of the outer disc and causes star formation to quench. 

In the right panel of Fig.~\ref{910_disk_fig}, we show the mean [Fe/H] and [Mg/Fe] abundances of the star-forming gas at each radius across the disc. The grey lines show the abundances in the CGM, computed edge-on with respect to the disc. Prior to the arrival of the Sagittarius analogue, there was a gradient in [Mg/Fe] across the disc. However, with the second approach of Sagittarius and coinciding with the onset of the disc tilting, [Mg/Fe] becomes nearly uniform across the disc. The accreted Sagittarius gas is generally more metal-poor and $\alpha$-rich than the disc (peaks in grey lines near the pericentres of Sagittarius), which likely contributes to metal dilution and $\alpha$ enhancement in the disc. The rising star formation rate during the interaction likely also contributes to the increasing [Mg/Fe] ratio in the centre of the MW analogue \citep{orkney}. 

Two effects are also noticeable in [Fe/H]. The first is that with the first two pericentres of the Sagittarius analogue, the metallicity in the inner disc is diluted and does not recover until the merger with the Sagittarius analogue is complete. This is consistent with the inflow of gas from the outer disc triggered by the Sagittarius analogue (reflected in the rising density in the inner disc and lowering density in the outer disc). The gas inflow from the outer disc also contributes to the flattening of the [Mg/Fe] gradient. The second is that after the GES-like merger, the disc was evolving at significantly higher metallicities than the CGM, while, following the second Sagittarius pericentre, the abundances in the outer disc follow closely the CGM abundances. The point at which the abundances of the outer disc and the CGM become comparable coincides with the disruption of the outer gas disc and therefore likely corresponds to the dilution of the outer disc metallicity with low-metallicity CGM gas \citep[as proposed by][]{orkney}.

\begin{figure*}
    \centering
    \includegraphics[width=1.6\columnwidth]{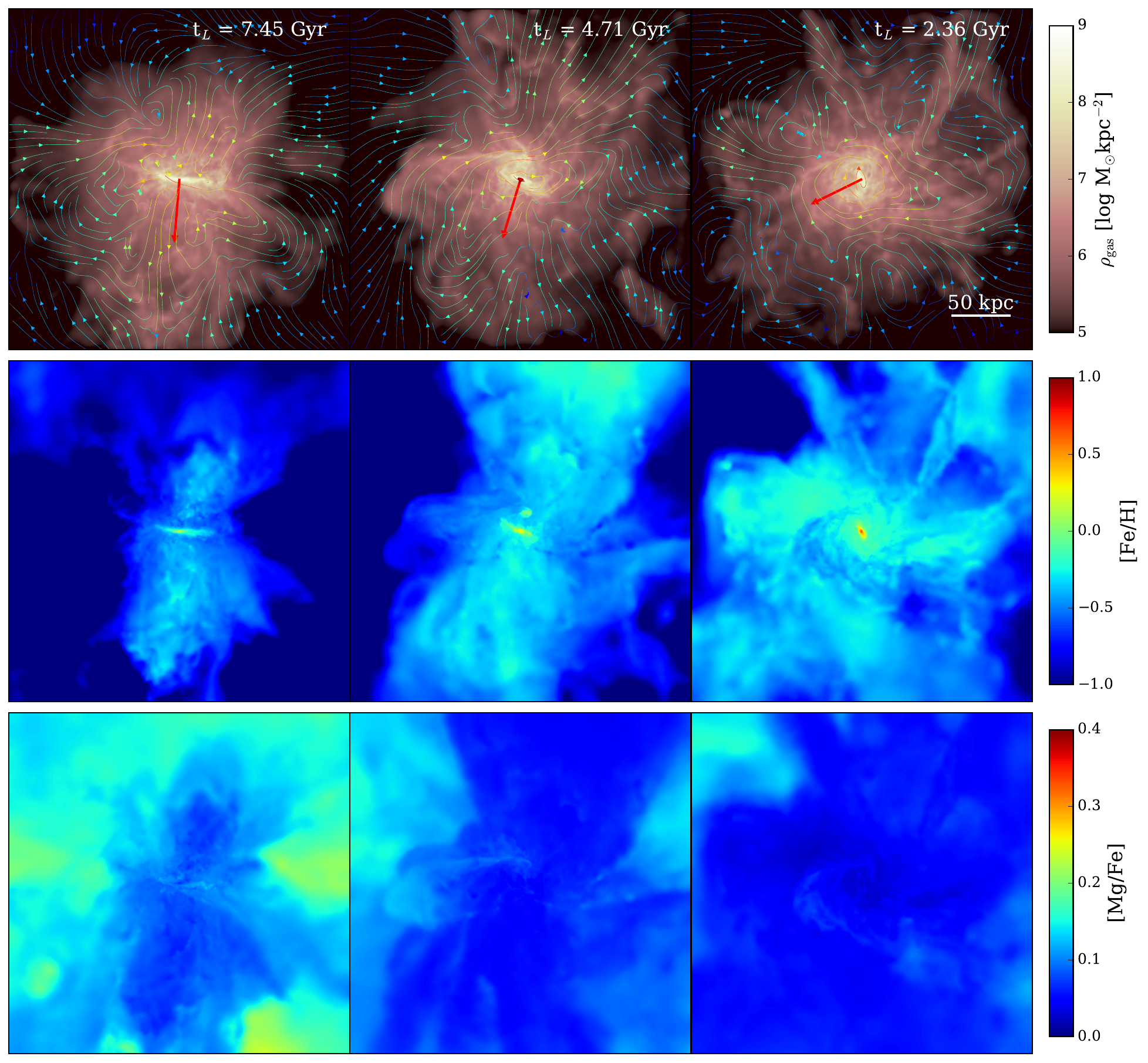}
    \caption{The evolution of gas density (top), [Fe/H] (middle) and [Mg/Fe] (bottom) for MW 910 during the arrival of Sagittarius-like analogue. The coloured arrows in the top row show the gas velocity field in a $\pm$50~kpc slice, with colours indicating the logarithm of the mass flux (the product of density and velocity magnitude). The thick red arrows show the direction of the angular momentum vector. }
    \label{fig:sag_outflow}
\end{figure*}

We also note that the spread in [Fe/H] and [Mg/Fe] in the CGM becomes smaller following the interaction with the Sagittarius analogue. We interpret this as the byproduct of disc tilting. This is demonstrated in Fig.~\ref{fig:sag_outflow}. Before the galaxy interacts with Sagittarius, the metals are concentrated in the disc and are expelled into the CGM through galactic outflows. The outflows, together with mostly metal-poor and $\alpha$-rich gas in the CGM return back along the disc plane. As the Sagittarius analogue approaches its second pericentre, the disc tilts, thus reorienting the metal-rich and $\alpha$-poor outflows (which are also enhanced in the interaction). As the disc tilts by $z$=0 by $\approx$70 degrees a larger fraction of the CGM is enriched by galactic metals, limiting the supply of $\alpha$-rich, metal-poor gas. Only in the aftermath of the interaction does some of the $\alpha$-rich gas return to the galaxy. 

Finally, we note a significant change in the gas metallicity and density around 2 Gyr ago. The change is accompanied by a large peak in star formation in the inner 10~kpc that coincides with the arrival of the LMC-SMC analogue pair and causes the formation of a large galactic bar in this galaxy. The formation of the bar increases the number of stars in the low-$\alpha$ sequence, although we note that the sequence is still clearly present if only stars older than 2~Gyr are considered (see Fig.~\ref{fig:910_by_age}).

The formation of a chemical bimodality in MW 910 is reminiscent of the mechanism proposed by \citet{grand_discs}, where consumption of gas in the outer disc leads to a pause in star formation, producing an $\alpha$-gap between the two sequences. In our simulations, the low-$\alpha$ sequence forms through the interaction with a retrograde Sagittarius analogue, which lowers the gas density in the outer disc (reducing the stellar population between the low-[Fe/H] ends of the two sequences) and drives gas inflow from the outer to the inner disc and increasing star formation rate, leading to a flat gradient in [Mg/Fe].

\citet{agertz} and \citet{vintergatan3} describe a path to the formation of a chemical bimodality, whereby a gas-rich satellite contributes gas in the formation of a misaligned metal-poor disc, which later realigns with the galaxy and is responsible for the formation of the metal-poor end of the low-$\alpha$ sequence. We have verified that the angular momentum vectors of the $\alpha$-rich and $\alpha$-poor sequences (as well as stars of different bins in age) stay well aligned throughout, suggesting that the mechanism presented here is distinct from the one described in \citet{agertz} and \citet{vintergatan3}.

\begin{figure*}
    \centering
    \includegraphics[width=2\columnwidth]{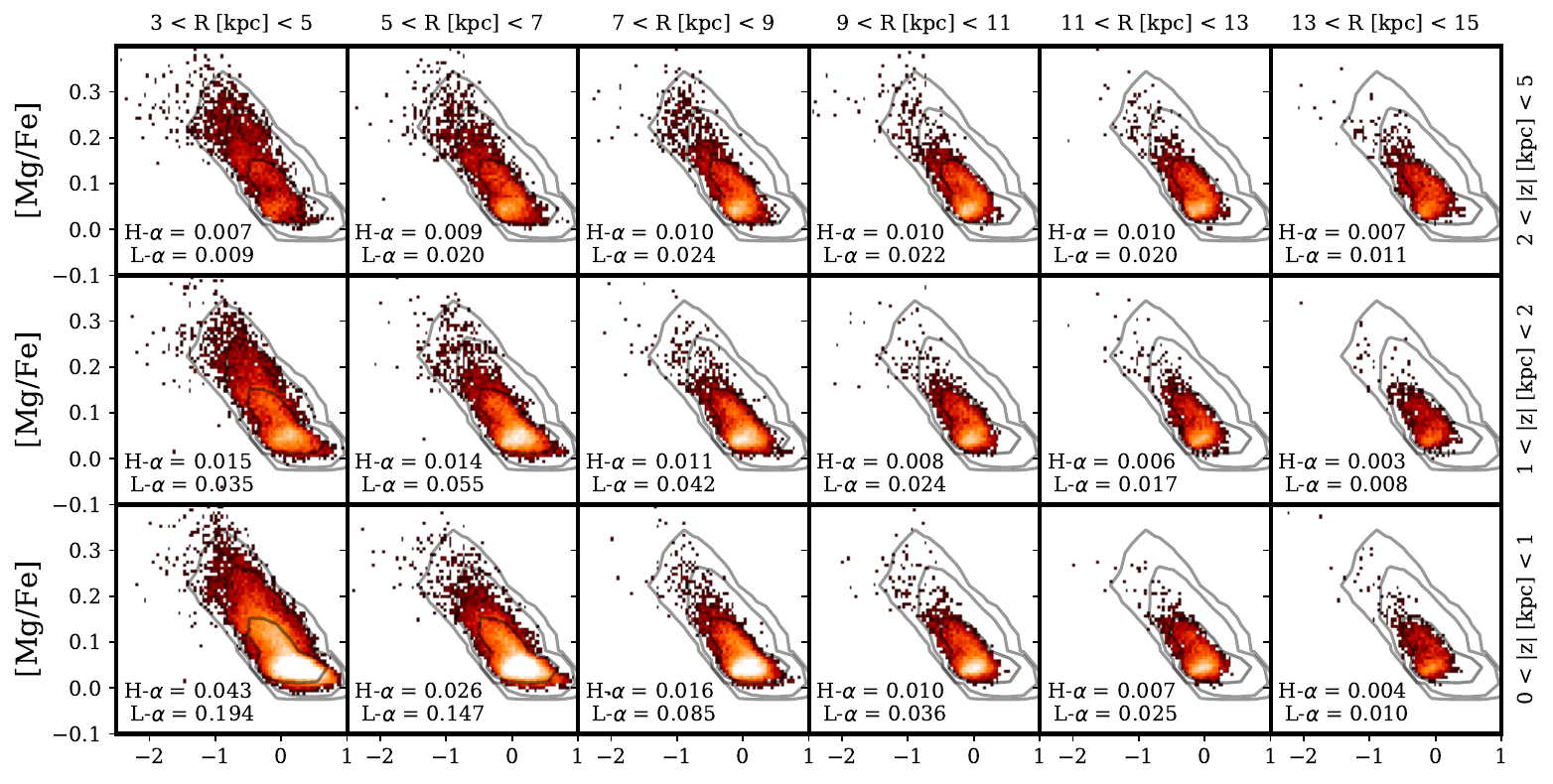}
    \caption{The [Mg/Fe]-[Fe/H] distribution of stars in MW 910 for 2~kpc bins in radius from 3 to 15~kpc (indicated at the top) and 3 bins in vertical height (indicated on the right). The colours are normalised by the total mass distribution across all radial and height bins. The grey contours show the 68$^{\rm th}$, 95$^{\rm th}$ and 99$^{\rm th}$ percentiles of the total distribution of \textit{in-situ} stars with $|z| < 5$~kpc as in Fig.~\ref{smooth_gas_props}. The high and low-$\alpha$ sequences are defined within the innermost contour, with the [Mg/Fe] cut as in the right panel of Fig.~\ref{fig_8_kpc}. We also quote the fractions of low and high-$\alpha$ stars in each bin, with respect to all stars across all bins in radius and height.}
    \label{fig:910_by_radius}
\end{figure*}

In Fig.~\ref{fig:910_by_radius}, we display the [Mg/Fe] and [Fe/H] abundances in concentric shells in the plane of MW 910's disc, with three vertical cuts, as in e.g. \citet{hayden}. The grey contour shows the distribution of the overall population, as in Fig.~\ref{fig:chemistry}. The high-$\alpha$ sequence stars are present over the whole radial extent of the disc. The locus of low-$\alpha$ sequence, however, changes across the disc, with high-metallicity stars in the central regions of the galaxy and low-metallicity stars primarily in the outer regions, with the bimodality being the most distinctive between 5 and 9~kpc. The low-$\alpha$ sequence in this galaxy is in good agreement with observational data for the Milky Way \citep{imig} and reflects the inside-out growth of the thin disc that must have followed the GES merger. The high-$\alpha$ sequence is however present in the outermost radii for MW 910, suggesting that its high-$\alpha$ disc is more extended than in the Milky Way \citep[e.g.][]{imig}.

\begin{figure*}
\begin{multicols}{2}
\includegraphics[width = \columnwidth]{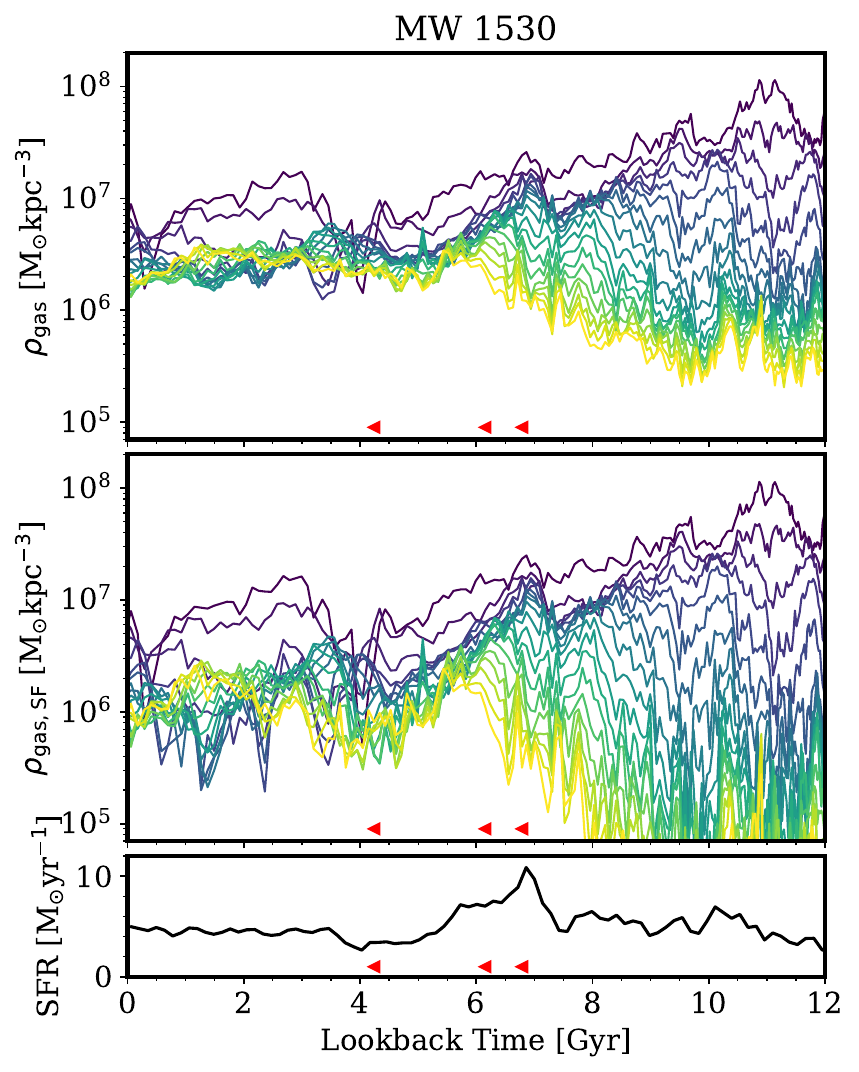}\\
\includegraphics[width = \columnwidth]{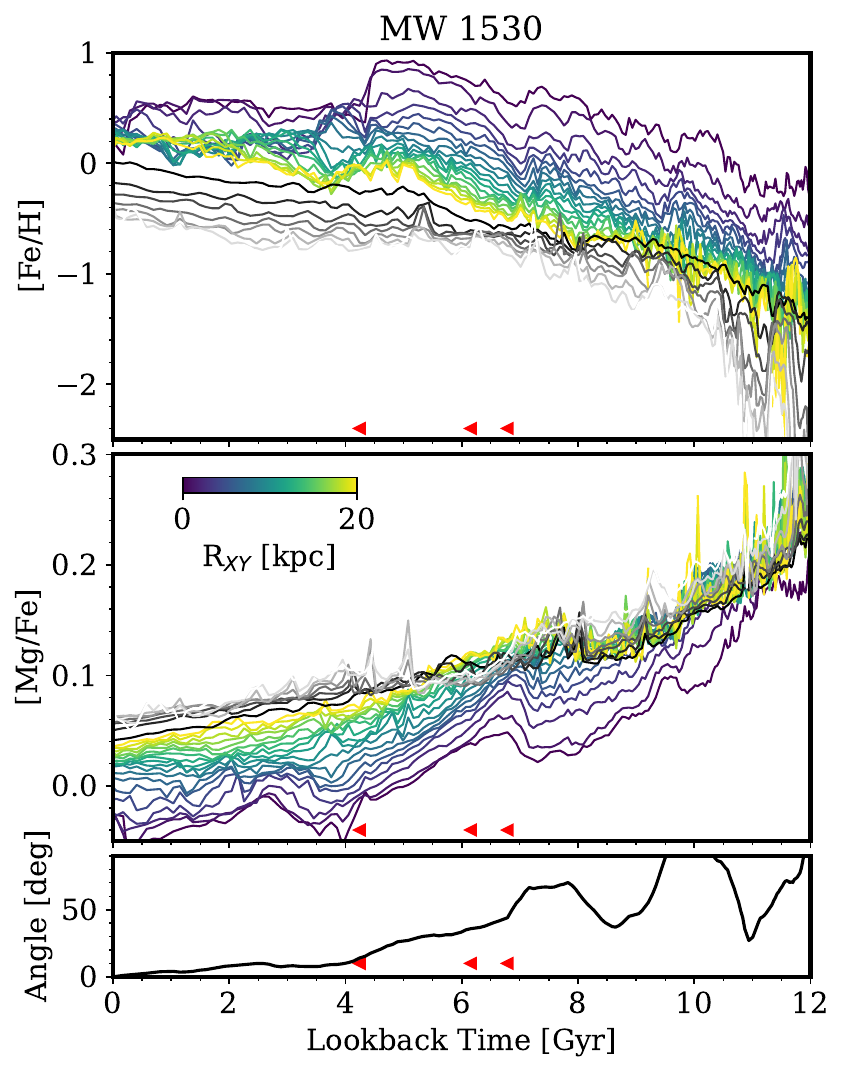}

\end{multicols}
\caption{Same as Fig.~\ref{910_disk_fig}, but for MW 1530.  }
\label{1530_disk_fig}
\end{figure*}

\subsubsection{Origin of bimodality in MW 1530}

In Fig.~\ref{1530_disk_fig}, we show the formation of the two sequences following a series of prograde mergers in MW 1530. After two mergers $\sim$7~Gyr ago with gas mass ratios 2.5~:~1 and 10~:~1 (dark matter ratio 5~:~1 and 10~:~1), the metal content is diluted (4.5~Gyr ago), and a low-metallicity end of the high-$\alpha$ sequence forms. Following a massive starburst starting before the coalescence with the first merger, we see the depletion of gas while the metallicity is rising across the disc and [Mg/Fe] is lowered. After coalescence during another merger 4~Gyr ago (gas mass ratio 4~:~1, dark matter 5~:~1), much of the gas is evacuated from the central regions, while the outer disc resumes growth soon thereafter at high [Fe/H]. We note the high [Fe/H] with a shallow [Fe/H] gradient across the low-$\alpha$ disc, that leads to the $\alpha$-rich and $\alpha$-poor sequences overlapping only over a narrow metallicity range for most radii. One can also see that the central part of the low-$\alpha$ disc (< 3~kpc) has a higher metallicity than the remainder of the disc, and forms a third high-[Fe/H], low-$\alpha$ population, as can be seen in the marginal histogram of Fig.~\ref{fig:chemistry}.  We also note that, unlike MW 910 in the last 5 Gyr of evolution, the metallicity in the CGM is significantly offset from the disc metallicity, suggesting that dilution from CGM gas may be important, especially in the outskirts of the galaxy. As seen in Fig.~\ref{fig:chemistry}, MW 1530 forms stars at lower [Fe/H] and higher $\alpha$ than MW 910 at late times, leading to a steeper slope in [Mg/Fe] of the $\alpha$-poor sequence. Overall, this pathway is consistent with the ``central starburst'' mechanism described for {\sc Auriga} galaxies in \citet{grand_discs}, although the star formation rate drops across the whole galaxy in this case.

The significant mergers experienced by MW 1530 are all prograde and thus tend to raise the density of gas, including star-forming gas, in the outer disc, with any reduction in gas density occurring primarily due to gas consumption in the merger-induced starburst. This density increase is in contrast to the retrograde merger in MW 910, where prior to the merger the gas density is significantly reduced and builds up slowly after the merger is complete. We observed the reduction in the gas density of the outer disc also for the retrograde merger in MW 720 (6~Gyr ago, see Fig.~\ref{fig:mass-growth}) and the increase in the outer disc density for the prograde merger 2~Gyr ago for MW 1540.

\subsubsection{The slope of the low-$\alpha$ sequence}

The last two sections reveal that there are multiple ways of producing a bimodality. The various mechanisms may actually be distinguishable from the slope of the $\alpha$-sequence \citep[see also][]{orkney,mason_slopes}. In MW 910, the slope in [Mg/Fe] of the low-$\alpha$ sequence is clearly flat, while in MW 1530 there is a clear negative gradient with metallicity.

These trends, as discussed previously, are readily explained by the composition of the gas in the disc and the inflows from the CGM in our simulations. In MW 910, the flat slope is set by the relatively uniform [Mg/Fe] composition in both the disc and the CGM. In MW 1530, [Mg/Fe] has a steep radial gradient within the disc, leading to a downwards trend overall. Locally (9-11~kpc radius), the locus of the low-$\alpha$ sequence in MW 1530 has a shallower gradient (left of Fig.~\ref{fig_8_kpc}) with an upturn at low metallicity as a result of dilution following the galaxy-wide starburst.

These results suggest that quantification of the slope differences (in both the high and low-$\alpha$ sequences) as well as where the two sequences join (e.g. Fig.10 of \citealt{imig}), in conjunction with stellar ages, will help narrow down their origin for the Milky Way. Such a quantification will clearly benefit from comparisons with galaxy formation and chemical evolution models.

\subsection{Some key points and the culprit merger}

Comparing MW 910 and MW 1530 serves to highlight some important criteria for the formation of chemical bimodality as observed in the Milky Way:

\textit{i)} The high-$\alpha$ sequence likely formed in a merger-induced starburst with a temporary reset in the metallicity of the disc, given its rapid evolution in [Fe/H] and $\alpha$-abundances and its resemblance to the high-$\alpha$ sequence in MW 1530 (Fig.~\ref{fig_8_kpc}).

\textit{ii)} Some process in the Galaxy must have prevented further enrichment to a higher [Fe/H] and lower-$\alpha$ after the starburst (i.e. when the low-$\alpha$ sequence started forming). In MW 910, this is the result of gas inflow from the Sagittarius analogue as well as the inflow from the outer disc induced by its retrograde orbit. At late times, iron enrichment is also stalled due to the formation of the galactic bar. In MW 1530, the starburst consumes the metal-rich gas from the central regions, allowing more metal-poor gas to flow in and reset the metallicity.

\textit{iii)} The Galaxy must have been able to achieve successful dilution at large radii to form the low-metallicity end of the low-$\alpha$ sequence, suggesting that the outer disc should not be significantly dense in metal-rich gas, before the inflow of metal-poor gas. In MW 910 this is achieved by the ``shrinking'' of the outer disc induced by the retrograde merger. In MW 1530 the gas is temporarily depleted following a galaxy-wide starburst, allowing more effective metal dilution of the gas in the outer disc, although this effect is short-lived as this galaxy soon regains metal-rich gas in the outer disc through accretion from satellites and further enrichment through star formation. 

\textit{iv)} [$\alpha$/Fe] must be to a large extent uniform and low across the disc at the time the low-$\alpha$ sequence forms. This likely requires efficient metal mixing across the disc. The low [$\alpha$/Fe] could be achieved either through Fe enrichment following a starburst and a decline in the star formation rate across the disc, as in MW 1530, and / or if the accreted gas in the formation of the low-$\alpha$ sequence is already at low [$\alpha$/Fe] due to being pre-enriched by the Galaxy. The latter, in particular, helps to maintain a shallow slope in [$\alpha$/Fe] -- [Fe/H] plane of the low-$\alpha$ sequence, as in MW 910.

Interestingly, \citet{funakoshi} find evidence of shrinking of the Milky Way's disc in the APOGEE and {\it Gaia} data and suggest that this is the result of gas consumption following the GES merger with delayed regrowth due to the transition from the cold to hot-mode accretion. Firstly, in our MW analogues hot-mode accretion is already dominant by the time of the GES merger (approx. 10 Gyr ago). Secondly, while rapid consumption of gas across the disc following a merger (as hypothesized by \citealt{funakoshi}) occurs in MW 1530, this is the result of a 4-body interaction (2 gas-rich mergers followed by a third merger) 7~Gyr ago, when the stellar disc is already established with high surface density. Such an interaction would likely result in stronger and more spatially extended gravitational instabilities throughout the disc, leading to a more significant starburst in MW 1530 compared to an early merger in MW 910 \citep{renaud_starburst}. Moreover, the gas fraction in MW 1530 drops rapidly from $f_{\rm gas}=0.14$ to 0.11 as a result of star formation and strong AGN feedback, resulting in a more intense and prolonged starburst \citep{cox, hopkins_mergers, fensch_mergers}, compared to an essentially unchanged gas fraction ($f_{\rm gas}=0.14$) in MW 910\footnote{We also note that generally prograde mergers in our simulations tend to increase the gas density in the outskirts of the disc (e.g. GES analogue merger in MW 910). }. It is thus unclear whether the GES merger could have triggered a starburst strong enough to shrink the outer gas disc of the Milky Way, or deplete it of the star-forming gas, while disc shrinking would be a likely outcome of a retrograde merger. 

Currently, the Milky Way is accreting the Sagittarius dwarf galaxy, and there is suggestive evidence of its impact on the star formation history of the Galactic thin disc \citep{ruiz-lara}. As far as we know, its impact on the chemistry of the thin disc has not yet been explored. The Sagittarius analogue in MW 910 clearly has an effect on the star formation of our Milky Way analogue through shocks and gravitational torques and contributes to the gas reservoir in MW 910 through its own gas and induced smooth gas accretion. Perhaps this could explain the recently reported evidence for a "second episode of thick disc formation", around $\sim 6$ Gyr ago, embodied in a population of stars of [Fe/H]$\sim -0.6$ dex, at [$\alpha$/Fe]$\sim 0.2$ dex, and which have a scale-height of $\sim 0.6$ kpc \citep{lian_second_thick_disk}. Additionally, \citet{das_sag} find evidence of an impulsive past interaction of the Milky Way's disc with Sagittarius, causing an increase in the velocity dispersion of stars formed prior to the interaction, as we see in our simulations (Fig.~\ref{fig:sfr_disk}). 

Although MW 910 displays a merger history similar to that inferred for our Galaxy, it is unlikely that the "real" Sagittarius could have triggered a similar disc disruption and tilting as in our simulations. While the infall mass of Sagittarius could have been sufficiently dynamically significant ($\gtrsim6\times10^{10} M_{\odot}$, e.g. \citealt {gibbons_sag}), the ``shrinking'' and tilting of Milky Way's gas disc requires that the orbit of Sagittarius is retrograde, with intermediate inclination with respect to Milky Way's disc \citep{villalobos,dodge_tilting}. Sagittarius is, however, observed to be in a polar orbit around the Milky Way \citep{tango}, which is unlikely to cause significant disc tilting, while the kinematics of its stellar streams suggest prograde motion \citep{sag_prograde}. GES, on the other hand, has deposited stars with retrograde motions around the inner Galaxy \citep{koppelman_retro} 
and could also explain retrograde overdensities observed in the Milky Way's halo \citep{naidu_retro,chandra_echoes}. It is also hypothesied to have caused a tilt of the Galactic disc \citep{thomas_tilting}. This makes a retrograde GES a more likely culprit for triggering the formation of a chemical bimodality in the Milky Way.

On the other hand, through integrating the torque from the collisionless Sagittarius analogue particles in the direction perpendicular to the disc plane of MW 910, we estimate that this accounts for $\approx$20~degrees of the disc tilting between the satellite's infall and disruption, suggesting that some of the disc tilting may be the result of mass momentum transfer in gas accretion. The full mechanism behind the tilting of galactic discs, particularly in the presence of gas, is outside the scope of this work; though the existence of old $\alpha$-poor stars in the Milky Way suggests that the origin of this sequence is more likely tied to an earlier accretion event, such as the GES, although Sagittarius could have enhanced the formation of the $\alpha$-poor stars.

\section{Discussion}
\label{sec:discussion}

We now place our findings regarding the formation of the [Fe/H]--$\alpha$ bimodality in MW~910 and MW~1530 in context of the current understanding of the formation of our Galaxy. In Section~\ref{sec_smooth_discussion}, we further compare our results regarding the enhanced smooth accretion following a galaxy merger to previous studies and discuss their implications regarding the contribution of smoothly accreted gas to the formation of the Milky Way's low-$\alpha$ disc.

\subsection{The Milky Way context}
 
The results presented in the previous section help us to understand the presence of the [$\alpha$/Fe] bimodality in the Milky Way. The last significant merger of the Milky Way was with GES, approximately 8-10 Gyr ago. The interaction is likely to have triggered a significant burst in star formation, forming much of the thick disc \citep{snaith_thick_disk, gallart}, as in MW~910 and MW~1530. 

If the merger with GES was retrograde \citep[as suggested by][]{enceladus,naidu_retro}, the star formation in the outer disc would be substantially suppressed during the merger. As the star formation rate drops in the inner disc following the starburst and the outer disc due to disc shrinking, Type~Ia supernovae could lower the $\alpha$ abundances across the disc \citep[see also][]{johnson_retro,amarante}. Alternatively, as in MW~1530, a prograde merger could also have shrunk the disc through consumption of star-forming gas in a disc-wide starburst. The merger likely led to a significant accretion of metal-poor gas from two main sources: gas that was bound to the GES and the smoothly accreted gas (Fig.~\ref{fig:mass-growth}). The accreted gas could then fuel the regrowth of the disc and the formation of the thin disc at low $\alpha$. As demonstrated in \citet{dodge_tilting} and \citet{thomas_tilting}, the disc is likely to have undergone substantial tilting during and after the merger, with outflows enriching a large fraction of the CGM with Galactic metals and removing differences in [$\alpha$/Fe] and [Fe/H] between the CGM and the outskirts of the disc, resulting in a lesser effect of metal dilution by the inflowing gas from the CGM and the low [$\alpha$/Fe] scatter observed in the $\alpha$-poor sequence.

It is worth noting that our GES-analogue in MW~910 is significantly higher in stellar mass ($M_* = 4 \times 10^9 M_{\odot}$) than some chemo-kinematic analyses of the data suggest by almost an order of magnitude \citep{lane_gse_mass}, but not others \citep{vincenzo,feuliet}. The halo mass of $M_{\rm DM} = 1.2 \times 10^{11} M_{\odot}$ is consistent with inferences based on GES stellar kinematics \citep{das_gse_halo_mass}. 

In Fig.~\ref{sag_ges}, we show the abundance distribution of the stars of the GES-analogue, which are significantly higher-metallicity than observed, as expected from the stellar mass of the analogue.  The $\alpha$-abundances of the GES-analogue are consistent with the tip of the high-$\alpha$ population, and indeed the gas brought by the GES-analogue primarily contributes to the formation of the high-$\alpha$ sequence. If a similar process took place in the Milky Way, it might help explain why some of the highest metallicity stars in GES have similar [$\alpha$/Fe] abundances as the most metal-poor stars in the sequence formed after the merger (in the case of the Milky Way, then thin disc), as noted by \citet{vincenzo}.

\citet{barbolato} have reported a period of coevolution of the thin and thick discs, with some disc stars in the low-$\alpha$ sequence predating the GES merger. In \citet{borbolato2026}, the old low-$\alpha$ stars were also found in the "splash" component of the halo \citep{pdmatteo_splash, Belokurov2020}. Fig.~\ref{fig:sfr_disk} demonstrates that some of the low-$\alpha$ stars in MW 910 at 8~kpc are forming already before and during the GES-analogue merger. In Fig~\ref{fig:910_by_age}, we also show that the stellar particles corresponding to the $\alpha$-poor population in MW 910 form over a wide range of metallicities prior to the merger, with typically higher $\alpha$ compared to the average for the low-$\alpha$ sequence. While this scatter in the stellar abundances can explain the presence of low-$\alpha$ stars at high redshift, it would struggle to explain the presence of two sequences with age > 11~Gyr as claimed in \citet{barbolato}. In the period during the retrograde GES-like merger, the presence of two sequences with similar ages is easier to explain, if the star formation in the heated and diluted outer disc is ongoing during the merger, albeit at a low level, as in MW 910, although the age overlap (but not the gap in $\alpha$) between the two sequences is more pronounced for prograde rather than retrograde mergers, as shown in \citet{amarante}. In MW~1530, with a prolonged galaxy-wide starburst, we also observe primarily sequential formation of the two sequences. 

\citet{beraldo_e_silva} and \citet{amarante} argue that old low-$\alpha$ stars in the disc may also originate from rapidly enriching high-redshift star-forming clumps. The ISM model employed in {\sc auriga} does not permit the formation of such clumps \citep{sh03,burger} and we therefore cannot comment on their contribution to the low-$\alpha$ sequence; however Figs.~\ref{fig:sfr_disk} and~\ref{figc2} demonstrate that prior to the mergers that heat up the high-$\alpha$ sequence, old stars in our simulated galaxies have low vertical velocity dispersions\footnote{We have also verified that the discy morphology of the old stars is present in many cases prior to the merger.}, which in some cases are consistent with the velocity dispersion of the low-$\alpha$ disc. Although mergers would dynamically heat the old disc, a colder component would have survived in the remnant, containing between 15\% and 25\% of the disc mass \citep{villalobos,Villalobos-growth}. This means that if a lower-$\alpha$ population was present at the time of the merger (originating from e.g clumps in the old thin disc), some of it would end up in the ``splash" component and some in the cold low-$\alpha$ disc. 

\subsection{Smooth gas accretion associated with mergers}
\label{sec_smooth_discussion}

One of the key findings of this work is that the smooth accretion of metal-poor gas is often enhanced during merger and could provide fuel for the formation of the low-$\alpha$ sequence at low metallicities. We note that signs of this effect were seen in previous studies. For instance, \citet{chandra} analyse a Milky Way analogue from the {\sc TNG}-50 simulations and find that directly following a merger there is an additional subsequent 30~per~cent growth in the mass of the dark matter halo that cannot be accounted for by the merger alone. \citet{buck_merger_impact} study the impact of early massive mergers on the chemical evolution of galaxies like the Milky Way in NIHAO-UHD simulations. Their figure~1 demonstrates that the cumulative gas mass growth cannot often be accounted for by the gas contributed by the mergers alone. They speculate that the growth in the cold gas could also be due to gas accretion triggered from the circumgalactic medium due to tidal forces exerted by the merger. \citet{sparre_mergers} track the origin of gas making up the merger ``bridge'' as well as the gas that contributes to star formation following a major merger. They also found that the gas originating from outside of the halo can contribute significantly to the star formation following a merger. \citet{merrow}, using {\sc Auriga}, show that a GES-like merger leads to the accretion of gas from the outskirts of the halo and beyond, helping the subsequent growth of the disc of their Milky Way analogue (Au-18), which does not occur in the simulation without a GES-like merger. The merger also reorients the disc and leads to a significant chemical bimodality \citep{orkney}. 

The smooth accretion of gas associated with mergers works in combination with the well-understood merger-induced torques that draw gas from the outskirts of galactic discs into their centres \citep{negroponte, barnes_hernquist, mihos_hernquist} and dilute the metallicity content \citep{rupke, montouri, bustamante_mergers}. In this work, we have shown that both the gas and the dark matter are drawn from beyond the virial radius of the galaxy during a merger. This gas will eventually form part of the circumgalactic medium and can be used in star formation subsequently. We also note that the enhanced smooth accretion of dark matter and gas during a merger is likely also accompanied by an enhanced rate of satellite accretion, as previously seen in \citet{dsouza_vell_focus}.

 As we have shown in Fig.~\ref{fig:mass-growth}, the smoothly accreted gas is more important in the early galaxy assembly, which is consistent with the results of \citet{mackereth_2018} for bimodal Milky Way-mass galaxies in the EAGLE simulations. Moreover, as \citet{Wempe2025} have shown, the Milky Way haloes in these constrained Local Group simulations assemble early compared to the general sample of Milky Way-mass galaxies in large cosmological volumes, making the early build-up of metal-poor gas reservoirs more likely. The early halo assembly may also imply the early formation of the Milky Way's disc \citep{aurora,conroy,rix,xiang_rix}, as demonstrated in \citet{semenov}.

In summary, the enhanced smooth accretion of gas associated with mergers is likely to occur in all simulations and galaxy formation models. Previous studies describe metal dilution from the ``metal-poor gas drawn from the CGM'' \citep{orkney, fire_bimodality}. Here, we attribute the presence of this gas to its enhanced accretion during mergers, thus making it available subsequently for star formation with lower metallicity, along with the typically more metal-poor gas contributed by satellites and mergers.

There is evidence that the Milky Way experienced significant mergers before Gaia-Enceladus \citep{kraken,horta2021,massari2026} and so the results presented here could imply hidden chemical sequences among stars in the hot thick disc/splash component \citep{enceladus,Belokurov2020}. However, these may be harder to disentangle, as can be seen from Fig.~\ref{smooth_gas_props}, which reveals significant overlap, and a much larger scatter in [$\alpha$/Fe] at low metallicity, reflecting stochasticity in galaxy formation for low-mass haloes at high redshift.

\section{Conclusions}
\label{sec:conclusions}

In this work, we have used a new suite of constrained-realisation cosmological MHD simulations of the Local Group, run with the {\sc Auriga} galaxy formation model. Our simulations use initial conditions in $\Lambda$CDM generated by \citet{Wempe2024} and produce Milky Way and M31 analogues with masses, separations and the local Hubble flow consistent with observations. The four Milky Way analogues in our suite are all discy and have disc and halo masses broadly consistent with observational constraints. We have focused in this paper on the formation of their stellar discs and, in particular, the chemical sequences in the $\alpha$- and Fe-elements and the presence of bimodalities.

We find in all our MW analogues that the gas delivered by mergers, gas enriched by mergers prior to accretion onto the halo and the smoothly accreted gas contribute nearly equally to the gas reservoir. Smoothly accreted gas generally dominates at early times (before 9-10 Gyr ago) since all our galaxies assemble quickly and in much denser environments during this time.

We find that merger and satellite infall is associated with increased smooth accretion of gas. The smoothly accreted gas was never bound to a halo prior to accretion onto the Milky Way and is thus unlikely to have been significantly metal-enriched. It also has higher $\alpha$-element abundance than satellite gas. The smoothly accreted gas further dilutes the metallicity of the CGM and allows the formation of more metal-poor stars in the following episodes of star formation. The low-metallicity smoothly accreted gas is particularly important at early times (lookback times $t_L >$ 12 Gyr), when gas is accreted to a large extent through cold flows. As a result, this gas is a primary contributor to some of the most $\alpha$-rich stars in the galaxy. As the mass of the halo increases, hot-mode accretion dominates, and the gas from which most stars form has spent significant time getting enriched in the halo. The transition to hot-mode accretion occurs in our simulations long before the mergers responsible for the formation of a chemical bimodality (occurring 12.6~Gyr ago for MW~910), this may be due to the earlier assembly of our Milky Way analogues \citep{Wempe2025}.

All four MW analogues show multiple distinct chemical sequences, but only two show clear [$\alpha$/Fe] bimodality. As in previous work with {\sc Auriga} \citep{grand_discs, grand_sausage, orkney}, we find that one pathway to form two $\alpha$-sequences is through a merger-induced burst of star formation, followed by a significant decline in star formation, allowing the new generation of stars to form at lower $\alpha$ abundances; however, the starburst must quench star formation over all the radii at which the bimodality is observed, to ensure a gap between the two sequences.

One MW analogue from our suite of simulations, MW 910, has a remarkably similar merger history as inferred for the Milky Way: a merger with a GES like object, an extended interaction with a Sagittarius-like dwarf galaxy and an LMC and SMC pair. The \textit{in-situ} stars in this simulation also show an [$\alpha$/Fe] bimodality very reminiscent in its pattern to that of the Milky Way, in the sense of the relative amplitude, slope and the location where the two sequences merge. In this simulation, the high-$\alpha$ sequence forms during and following a merger with a GES-like object, possibly over too extended a period compared to the Milky Way. The low-$\alpha$ sequence arises as a result of interaction with a Sagittarius-like satellite on a retrograde orbit. The interaction induces the inflow of more metal poor gas to the centre of the galaxy from the outskirts of the disc and from the satellite itself (preventing further enrichment at the high-Fe end) and disrupts the angular momentum in the outer disc, leading to quenched star formation in the outer disc and an $\alpha$ gap at low metallicity between the two sequences. However, we point out that the real Sagittarius is unlikely to have caused this in the Milky Way due to its polar and slightly prograde orbit. Instead, as previously investigated in \citet{amarante}, a retrograde GES merger could equivalently deliver this mechanism.

Another analogue, MW~1530, forms an $\alpha$--Fe bimodality through a merger-induced galaxy-wide starburst, which also results in the temporary quenching of the outer disc and is followed by the dilution of the outer disc with metal-poor gas. This provides an alternative mechanism for the formation of a chemical bimodality in the Milky Way; however, MW~1530 is the only galaxy in our sample where a starburst is able to effectively quench the outer disc. The starburst happens at late times, when the surface density of the disc is high, and involves multiple merging objects. It is therefore unclear whether a sufficiently significant starburst like this could have occurred at early times in the Milky Way. Moreover, the mechanism preserves a steep [Mg/Fe] gradient across the disc, leading to a steeper slope of the low-$\alpha$ sequence in the [Mg/Fe] -- [Fe/H] plane. The low-metallicity end of the low-$\alpha$ sequence also forms quickly in this galaxy (over 1-2 Gyr), unlike in the Milky Way \citep{imig}. We therefore favour a retrograde merger scenario, as in MW~910, over a wide starburst mechanism.

An interaction with a massive object like GES on an inclined retrograde orbit has two additional positive consequences for the formation of a chemical bimodality with a wide [Mg/Fe] gap, a [Fe/H] overlap and a shallower [Mg/Fe]--[Fe/H] gradient in the low-$\alpha$ sequence. Firstly, it leads to a strong burst of star formation with the quenching of the outer disc\footnote{Instead, we observe that for the GES-like merger in MW 910, which is on a prograde orbit, the galaxy is not quenched sufficiently in our simulations, at least in comparison to the observations.}, with rapid evolution from high to low $\alpha$. Secondly, it would lead to the flattening of the radial [Mg/Fe] gradient in the disc and the tilting of the disc, with the latter having an additional effect of redirecting the supernovae and AGN outflows, enriching a larger fraction of the CGM and removing radial metallicity and $\alpha$-gradients within the CGM, thus putting the CGM at a rough chemical equilibrium with the disc outskirts. This reduces the effect of dilution in the later stages of the formation of the galaxy and would contribute, as in MW~910 to the low scatter in [Mg/Fe] in the formation of the low-$\alpha$ sequence. 

Although we have not investigated the effect of radial migration in this work \citep{schroenrich}, the bimodality formation pathways we describe here do not require radial migration, with the main factor being the quenching over radii where the bimodality is observed. Radial migration may, nevertheless, enhance the chemical bimodality.

In comparison to similar hydrodynamical suites of simulations from the literature, such as {\sc Auriga} \citep{grand_auriga} and FIRE-2 \citep{fire_bimodality}, it is interesting to note that all our MW analogues show multi-modalities in [Mg/Fe]--[Fe/H], as do all those systems in group environments modelled with FIRE-2 \citep{barry_fire}, 
but not all Auriga  \citep{orkney}, which are generally isolated systems. This possibly suggests an environmental effect, perhaps related to the more quiescent history/suppressed late-time growth of MW-analogues in groups \citep{Wempe2025}. A so far unexplored characteristic of most existing simulations with bi-modalities, is that they often depict chemical sequences that are nearly parallel, i.e. with similar slope, whereas the data may suggest a shallower slope for the low-$\alpha$-sequence compared to the high $\alpha$-sequence \citep[see e.g.][]{orkney}. Proper quantification is necessary (taking into account selection effects and variation with distance from the Galactic centre) as it could serve as a diagnostic to pin-down the processes by which the MW developed its bimodality, as well as for constraining galaxy formation models. It is worth noting that at face value no {\sc Auriga} analogue shows a bimodality where both sequences have slopes similar to those observed, and that in this respect, and despite the shortcomings described above, MW 910 may well be the closest analogue.

\section*{Acknowledgements}
\addcontentsline{toc}{section}{Acknowledgements}
AG would like to thank Annette Ferguson, Matt Orkney and Gail Zasowski for helpful discussions. AG is supported by UK Research and Innovation (UKRI) under the UK government’s Horizon Europe funding guarantee [grant number EP/Z534353/1]. AH acknowledges financial support from NWO through a Spinoza prize (SPI-78411). RJJG acknowledges support from an STFC Ernest Rutherford Fellowship (ST/W003643/1). This research was carried out using the High-Performance Computing resources of the FREYA cluster at the Max Planck Computing and Data Facility (MPCDF, https://www.mpcdf.mpg.de) in Garching, operated by the Max Planck Society (MPG). This research made use of {\sc numpy} \citep{numpy}, {\sc scipy} \citep{scipy}, {\sc matplotlib} \citep{matplotlib} and {\sc pySphviewer} \citep{alejandro_benitez_llambay_2015_21703}.


\section*{Data Availability}
The data underlying this article will be made available upon reasonable request to the corresponding author.



\bibliographystyle{mnras}
\bibliography{bibliography-ah} 




\appendix
\renewcommand\thefigure{A\arabic{figure}}    
\setcounter{figure}{0} 

\section{Properties of the Milky Way and M31 analogues}

\subsection{Deriving disc and bar parameters}
\label{disk_params}

To kinematically select the thin and thick discs, we construct an azimuthally averaged basis function expansion of the potential from the simulation snapshots using {\sc agama} \citep{agama}. The young stellar disc and gas component are modelled with azimuthal Fourier harmonic expansions, while the dark matter and old stellar populations are modelled with multipole expansions. Since 3 of the 4 Milky Way analogues contain a bar, before computing 
orbital circularities we first remove the stellar particles belonging to the bar. This is done following the method presented in \citet{algorry_bar} and \citet{frakgoudi_bar}. In short, after the stellar disc is aligned such that its angular momentum vector points along the $z$-axis, the Fourier modes of the surface density are computed for particles with a disc height $|z|<0.8$~kpc. The bar strength, $A_{2,\rm max}$ is defined as the the maximum value of the $m=2$ component in the central 10~kpc. The bar length is the radius where $A_{2}$ drops below 0.7$A_{2, \rm max}$ and the bar position angle is $\phi_{\rm bar} = 0.5\tan^{-1}(b_2/a_2)$. The bar particles were selected within $|z|<5$~kpc and along the position angle and within the bar length such that the projected density in each direction perpendicular to the bar in the $x$-$y$ plane remains above a threshold $7\times10^8M_{\odot}$kpc$^{-2}$. On visual inspection, this threshold successfully removed the bar-shaped overdensity in the simulated galaxies. After the bar particles were removed, the orbital circularities were computed and a mixture model was fitted to their distribution consisting of two skewed normal distribution components, representing the thin and thick disc (with the skewed Gaussian distribution truncated at the circularity of 1 for the thin disc), and up to 6 Gaussian components to represent the spheroid. The total number of spheroidal components was determined using the Akaike and Bayesian Information Criteria \citep{akaike, bayesian}. MW~1530 and MW~1540 preferred 3 spheroidal components, while MW~720 preferred 4 components and MW~910 preferred 5. The mass of the thin and thick disc, as well as the total mass in the spheroidal component were then estimated from the decomposition. Since the spheroidal component includes a mixture of bulge and accreted stars, we also quote the fraction of accreted stars, $f_{\rm acc}$, in Table~\ref{tab:main-props}.

For fitting the double-exponential profile, the particles were further refined to have a vertical height $|z|<5$~kpc, galactocentric distance in the plane $R > 5$~kpc (which removes particles within the bar length) and smaller than either 15~kpc or the location of disc warp. Furthermore, following \citet{minchev_fit} and \citet{bernaldez_fit}, $|z|< 1$~kpc and $|z| > 0.2$~kpc cuts were applied when fitting the radial and vertical scale densities, respectively.

Since the selection of particles belonging to the bar also removes a fraction of the disc and spheroidal components in the central regions of simulated galaxies, we additionally estimated the mass of these components in the region occupied by the bar. The thin and thick discs were modeled with the previously obtained double-exponential profiles. Each spheroidal component was fit from the bar length and out to 20~kpc with a broken-power-law model. We then Monte Carlo sampled $10^6$ particles of each component to obtain its mass in the region occupied by the bar. This additional mass is shown in brackets in Table~\ref{tab:main-props}. Since bars originate in disc instabilities \citep{athanassoula_barform}, the method likely overestimates the present-day densities of the components in the bar region (for MW 1540, for example, the estimated mass mildly exceeds the mass of the weak bar in the galaxy); however, the estimated masses should reflect those prior to bar formation.

\subsection{Comparison with {\sc Auriga}}
\label{auriga_comp_sec}

\begin{figure}
    \centering
    \includegraphics[width=0.9\columnwidth]{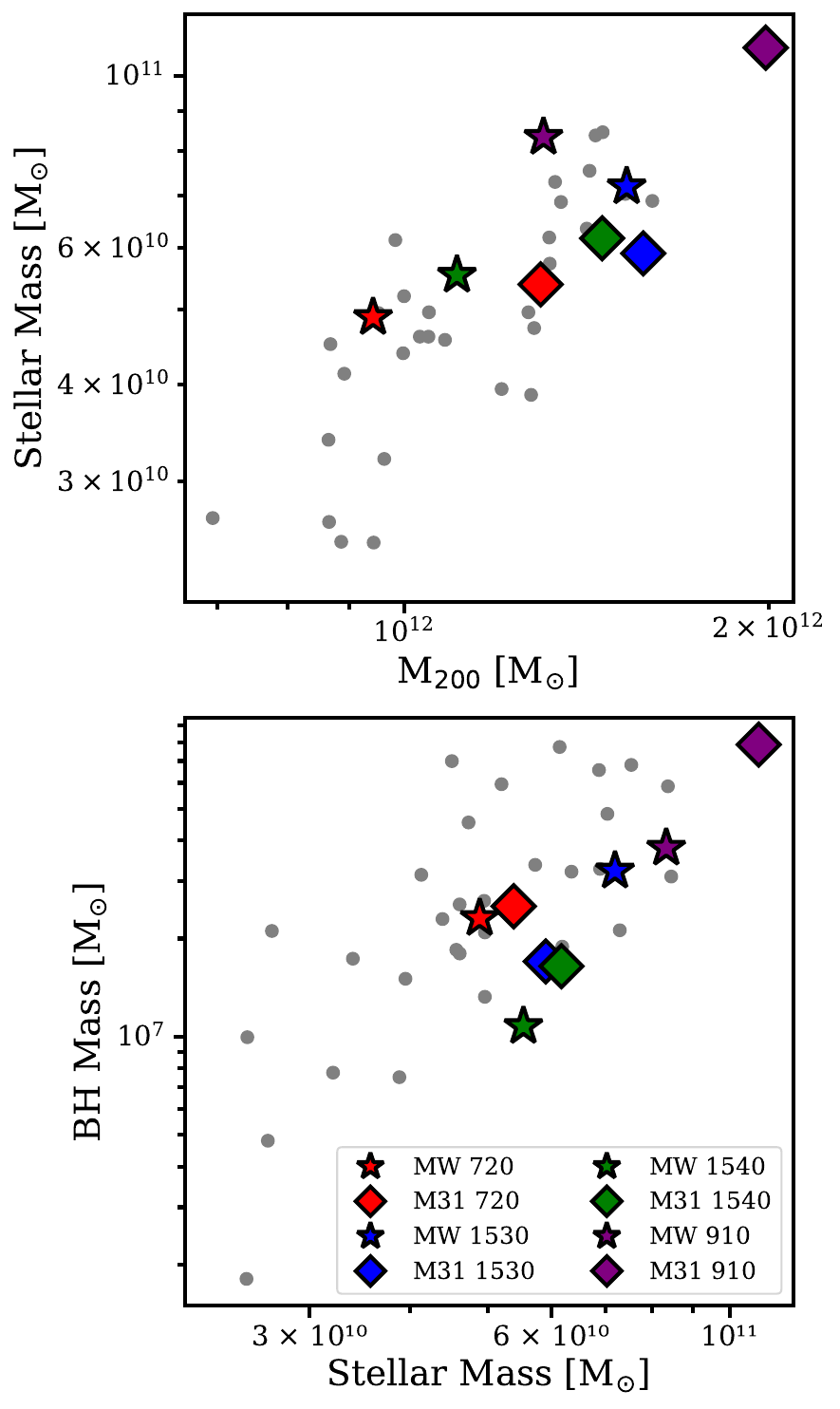}
    \caption{\textit{Top:} stellar mass within $R_{200}$ as a function of all mass within $R_{200}$. Particles and gas cells bound to the satellites of the central galaxy are excluded. The grey circles show 30 galaxies from {\sc Auriga} level-5. The colour symbols show the Milky Way (stars) and M31 (diamonds) analogues from our constrained simulations. \textit{Bottom:} the mass of the central black hole as a function of total stellar mass.}
    \label{fig:auriga_comparison}
\end{figure}

In Fig.~\ref{fig:auriga_comparison} we compare the stellar, total and black hole masses for our Milky Way and M31 analogues to similar-resolution central galaxies in {\sc Auriga} level-5. The stellar and halo masses are in good agreement with the full population of galaxies in {\sc Auriga}, aside perhaps from M31~910, which has a higher stellar and halo mass compared to the {\sc Auriga} sample. Our Milky Way analogues tend to have slightly higher stellar masses than the average for given halo mass in {\sc Auriga}, while the M31 analogues tend to have lower stellar masses. Our galaxies also tend to have slightly lower black hole masses than average. Nevertheless, a larger sample of Local Group analogues will be necessary to fully quantify any systematic differences from the isolated Milky Way and M31 analogues in {\sc Auriga}. We will aim to address these differences in future work.

\section{Properties of MW 910}

\renewcommand\thefigure{B\arabic{figure}}    
\setcounter{figure}{0}

\subsection{Cold vs. hot-mode accretion}
\label{app:cold-hot}

\begin{figure}
\centering
\includegraphics[width=0.8\columnwidth]{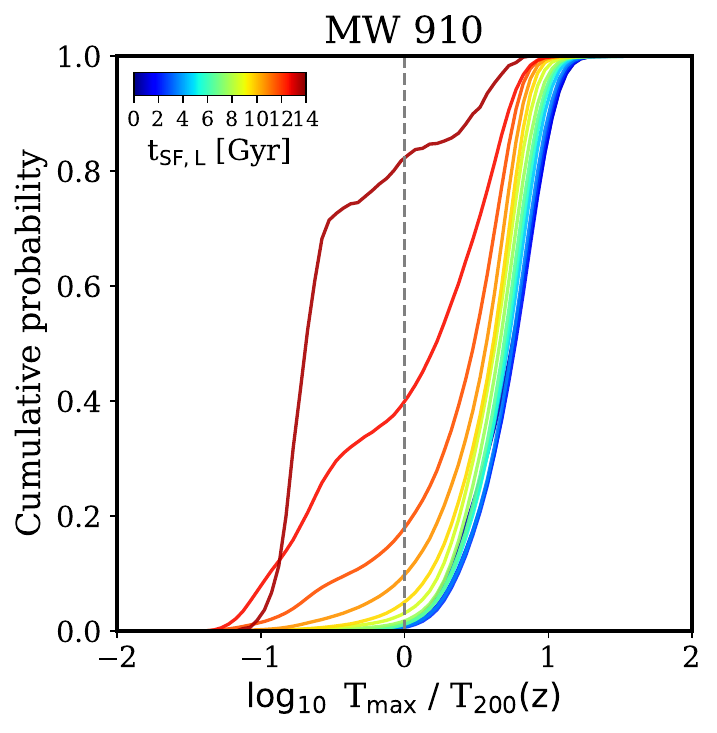}
\caption{The cumulative distribution of maximum temperature compared to the virial temperature of MW 910 for gas tracers that contribute to the stellar populations forming at each lookback time, as indicated by the colour bar. The virial temperature of the halo is taken at the time the tracer records its maximum temperature. The vertical grey shaded line indicates maximum temperature equal to the virial temperature. Maximum temperatures below this line indicate gas accretion through cold flows, while maximum temperatures above the line indicate hot-mode accretion.  }
\label{hotcold}
\end{figure}

In this section we investigate the importance of cold and hot-mode accretion in MW 910. We first split the stellar populations in the galaxy into bins of 1~Gyr of lookback time of formation. For stars in each bin, we identify the tracers that lie within the stellar population and track them back to their gas origin. For as long as the tracer has belonged to a gas cell we identify the maximum temperature that the gas had, excluding the times when the gas was star-forming (since for star-forming gas the temperature is raised artificially by the \citet{sh03} equation of state). We then compare this temperature to the virial temperature of the halo at that redshift, following \citet{keres}. The results are shown in Fig.~\ref{hotcold}. Cold-mode accretion dominates at the earliest times in the formation of the halo (with the transition from cold to hot-mode dominated accretion occurring approximately 12.6~Gyr ago) and becomes insignificant after the GES-like merger that completed 8~Gyr ago. It is therefore unlikely that cold flows would result in the metal dilution necessary to form a low-$\alpha$ sequence in our simulations as suggested in \citet{chiappini1,chiappini2}. This result is similar across the full range of MW analogues discussed in this work and is expected in moving-mesh simulations as opposed to smoothed-particle hydrodynamics treatments \citep{nelson_sph_arepo, nelson2016zooming}.

\subsection{Contribution of GES and Sagittarius gas}
\label{app:GES-Sgr-gas}

\begin{figure}
\centering
\includegraphics[width=0.8\columnwidth]{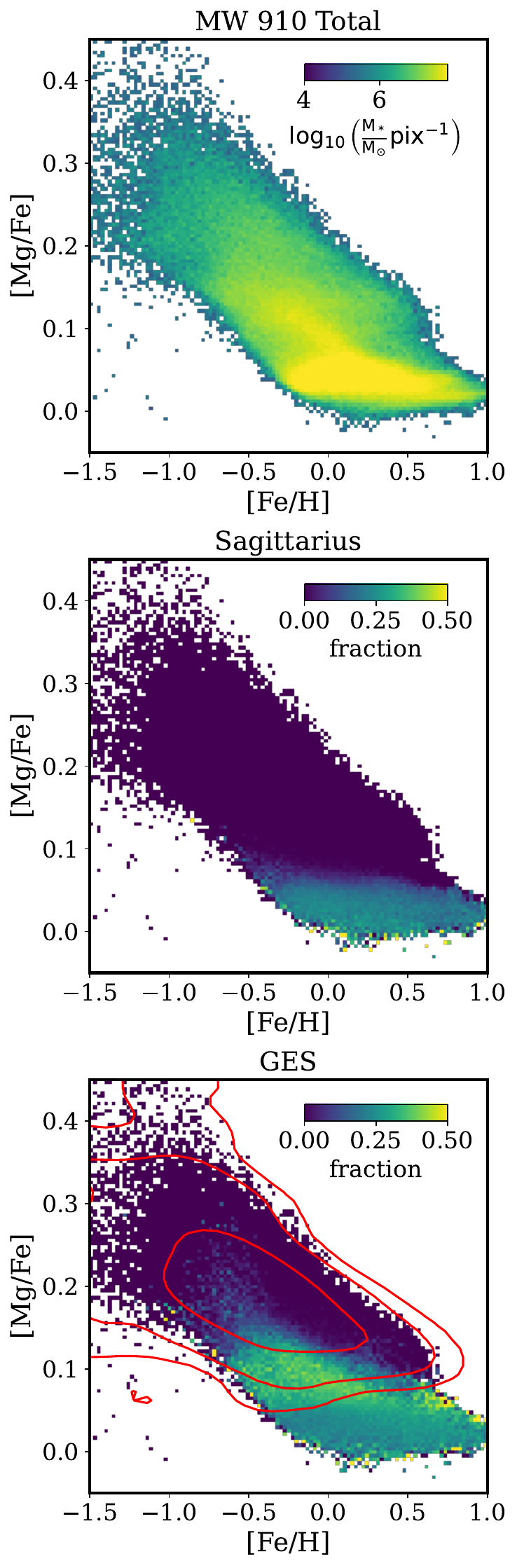}
\caption{Fractional contribution to the MW 910 \textit{in-situ} star formation by gas associated with the Sagittarius analogue (middle) and the GES analogue (bottom). The red contours in the bottom panel show the 68$^{\rm th}$, 95$^{\rm th}$ and 99$^{\rm th}$ percentiles of the accreted stars which were contributed by the GES analogue. Sagittarius analogue contributes $\sim$20~per~cent of the gas forming the $\alpha$-poor population, while GES analogue contributes $\sim$40~per~cent of the gas in the $\alpha$-rich population and provides a similar contribution as Sagittarius in the formation of the $\alpha$-poor population. Stars accreted from the GES analogue have on average higher $\alpha$ and lower [Fe/H] than the $\alpha$-rich population of MW 910. }
\label{sag_ges}
\end{figure}

We explore the contribution of the gas from the GES like merger and the Sagittarius analogue to the formation of the high and low-$\alpha$ sequences in MW 910, similarly to Fig.~\ref{smooth_gas_props}, but including the gas with past association to the two objects that was not present at the time of maximum gas mass. The top panel of Fig.~\ref{sag_ges} shows the overall mass distribution of insitu stars with $|z|<5$~kpc across the [Mg/Fe]--[Fe/H] plane, highlighting the location of the two sequences. The middle panel shows the contribution of Sagittarius gas, which makes up on average 20-30~per~cent of the stars in the low-$\alpha$ sequence. The bottom panel highlights the contribution of the gas from a GES-like merger, whic is close to half of the gas that made up the high-$\alpha$ sequence, as well as many of the stars with similar $\alpha$ abundances. It also provided a similar contribution as Sagittraius to the formation of the low-$\alpha$ sequence.

The red contours outline the chemical composition of the stars belonging to the GES-like merger, which is very similar to the distribution of Milky Way stars formed prior to the GES-like merger. As pointed out in \citet{orkney}, the chemical composition of stars formed from this gas is unlikely to bear signatures of the GES chemistry, as the gas is mixed with that of the Milky Way analogue and typically undergoes several enrichment episodes before forming stars.

\subsection{[Mg/Fe]-[Fe/H] distribution of MW 910 split by stellar age}

\begin{figure*}
    \centering
    \includegraphics[width=0.7\linewidth]{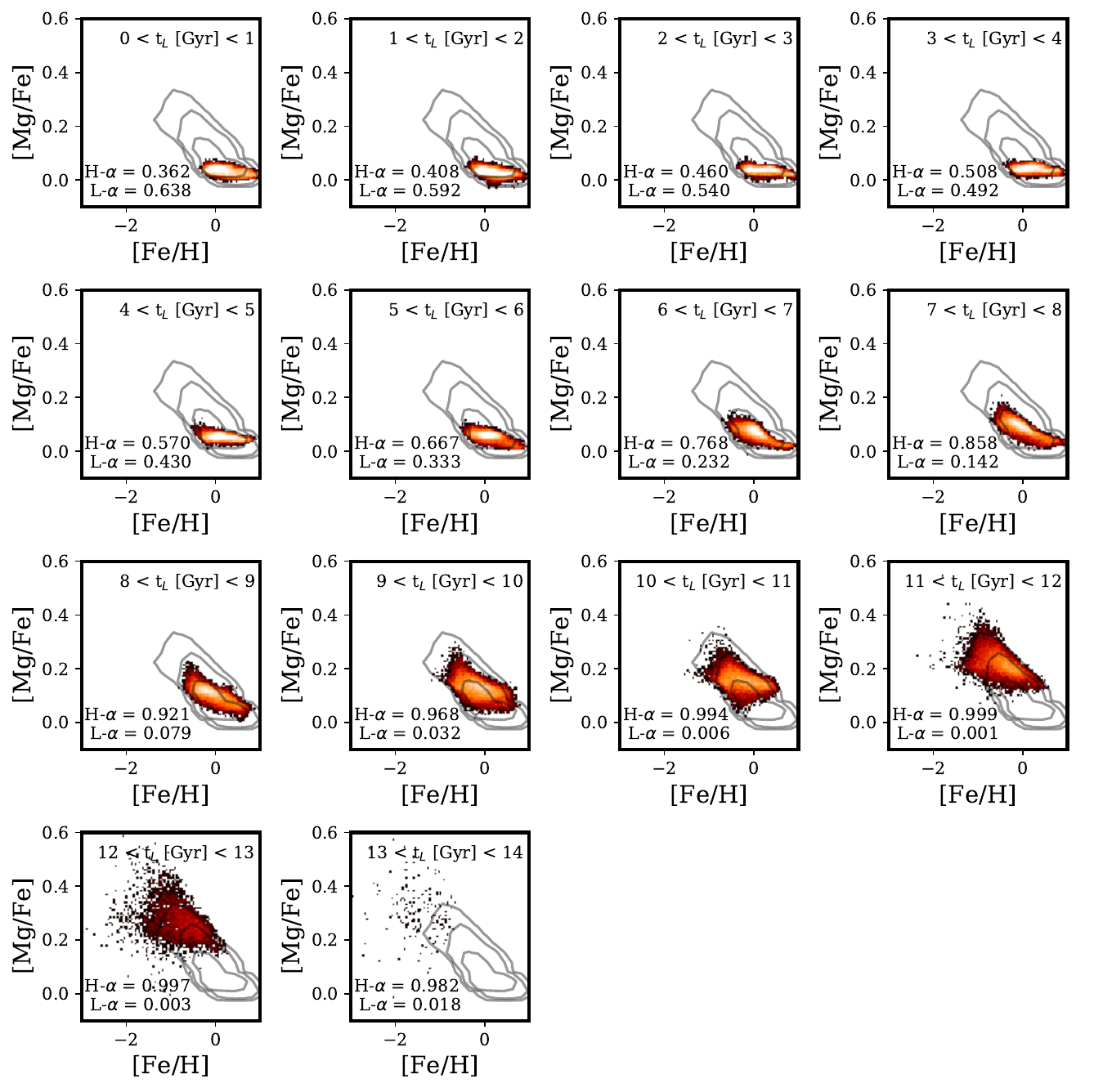}
    \caption{The [Mg/Fe]-[Fe/H] distribution of all \textit{insitu} stars  with $|z|<5$~kpc formed in each 1~Gyr bin of age (lookback time of formation) in MW 910. The colours are normalised by the integral of the total distribution at the present day. The labels show the fraction of all stars formed up to and including each age bin made up by the high and low-$\alpha$ stars.}
    \label{fig:910_by_age}
\end{figure*}

We explore how the [Mg/Fe]-[Fe/H] distribution is populated as a function of stellar formation time. Fig.~\ref{fig:910_by_age} shows the chemical composition of stars formed in each 1~Gyr bin of age (lookback time of formation). Earliest star formation consists of stars formed primarily from cold-flow gas, with lower metallicity in the outskirts of the galaxy, and higher metallicity in the central parts (see Fig.~\ref{smooth_gas_props}). Enrichment within the disc leads to the formation of supersolar metallicity high-$\alpha$ stars. After the GES merger, however, the locus is offset, defining and populating the high-$\alpha$ sequence. After 6~Gyr, we see a transition from a diagonal high-$\alpha$ sequence to a flat low-$\alpha$ sequence, following the interaction with Sagittarius.  We also see that some low-metallicity high-$\alpha$ stars form in the last 3~Gyr as more $\alpha$-rich gas is accreted with the arrival of the Magellanic Cloud analogues.

\subsection{Resolution dependence}
\label{resolution_910_sec}

\begin{figure}
    \centering
    \includegraphics[width=\columnwidth]{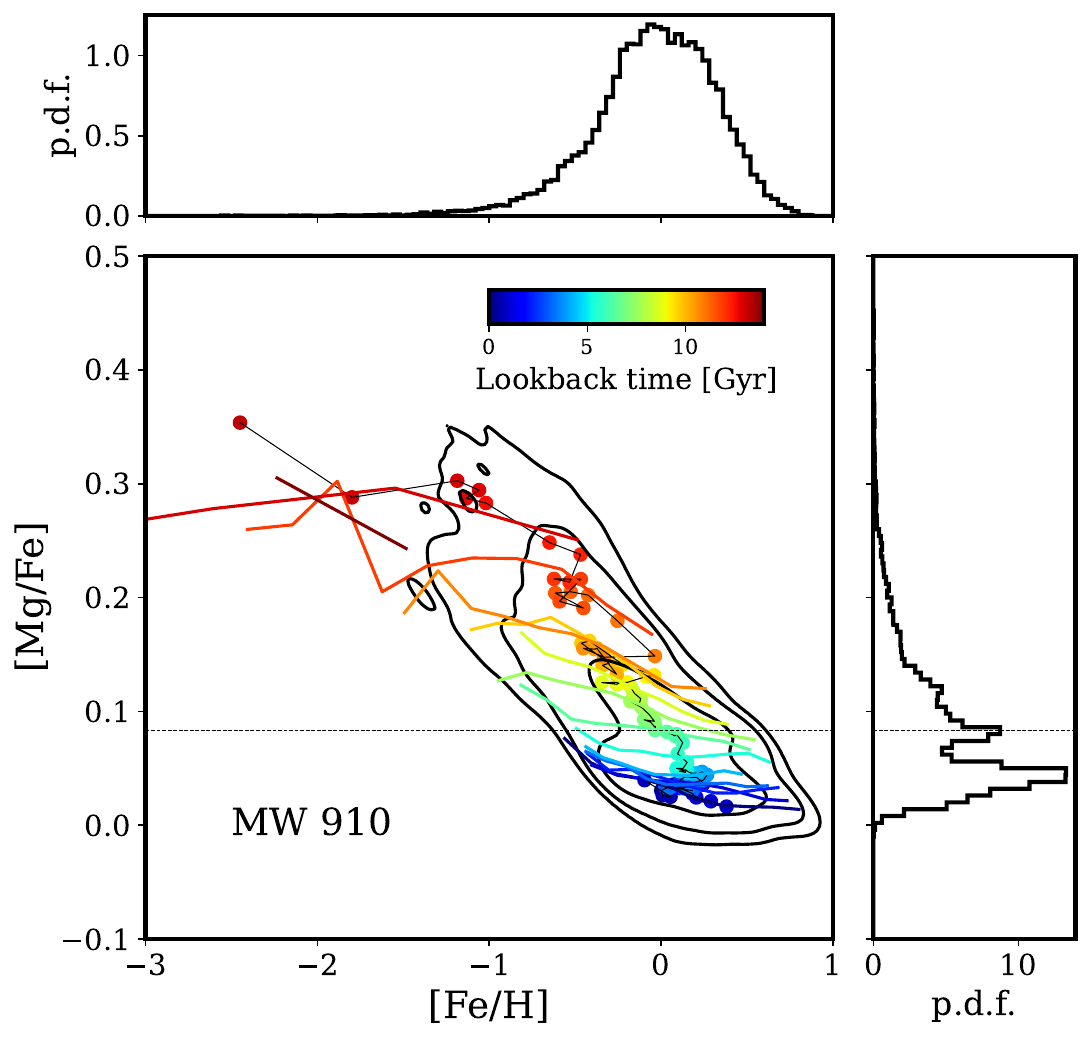}
    \caption{As Fig.~\ref{fig:chemistry}, but for lower-resolution ICs of MW 910. The black dotted line shows the same cut between the $\alpha$-rich and $\alpha$-poor populations of MW 910 that we obtained in the high-resolution simulations.}
    \label{fig:910_low_res}
\end{figure}

In addition to the simulations presented in this work, we have also run lower-resolution resolution simulations of the Local Group analogues, with the dark matter particle mass of $1.6\times10^7 M_{\odot}$. In Fig.~\ref{fig:910_low_res}, we show the [Mg/Fe]-[Fe/H] diagram, similar to Fig.~\ref{fig:chemistry}, for the lower-resolution realisation of MW 910. It can be seen that the properties are broadly consistent across resolution, showing the presence of a slanted high-$\alpha$ sequence and a comparatively flat and more prominent low-$\alpha$ sequence. There is a more pronounced drop in [Mg/Fe] during the period of interaction with the Sagittarius analogue, suggesting a more significant period of quenching compared to the higher-resolution realisation. This is associated with a peak in star formation caused by the interaction with the Sagittarius analogue $\approx6$~Gyr ago, which is in fact more significant than the peak caused by the GES-like merger for this realization.  

\section{Bimodality in M31 720}
\label{m31_720_sec}

One of the M31 analogues in our simulations, M31 720, also exhibited a clear bimodality in [Mg/Fe] - [Fe/H], resembling that of the Milky Way, which we briefly discuss here. Fig.~\ref{figc1} demonstrates the elemental abundances in this galaxy. M31 720 underwent a major merger (1~:~0.9 mass ratio) that completed 7.5~Gyr ago, leading to a significant burst in star formation and the formation of the high-$\alpha$ sequence (Fig.~\ref{figc2}). Following the burst in star formation and a declining SFR, [$\alpha$/Fe] reduced substantially until approximately 6~Gyr ago, when the accretion of metal-poor and $\alpha$-poor gas onto the disc led to the formation of the low-[Fe/H] end of the low-$\alpha$ sequence, after which the galaxy continued enrichment to higher [Fe/H] and lower [Mg/Fe]. Since the merger was prograde, the star formation rate in the outer regions of the galaxy did not change significantly during the merger, leading to a higher scatter in [Mg/Fe] and a smaller gap in [Mg/Fe] between the two sequences at low [Fe/H], compared to MW 910.

\renewcommand\thefigure{C\arabic{figure}}    
\setcounter{figure}{0}

\begin{figure}
    \centering
    \includegraphics[width=\columnwidth]{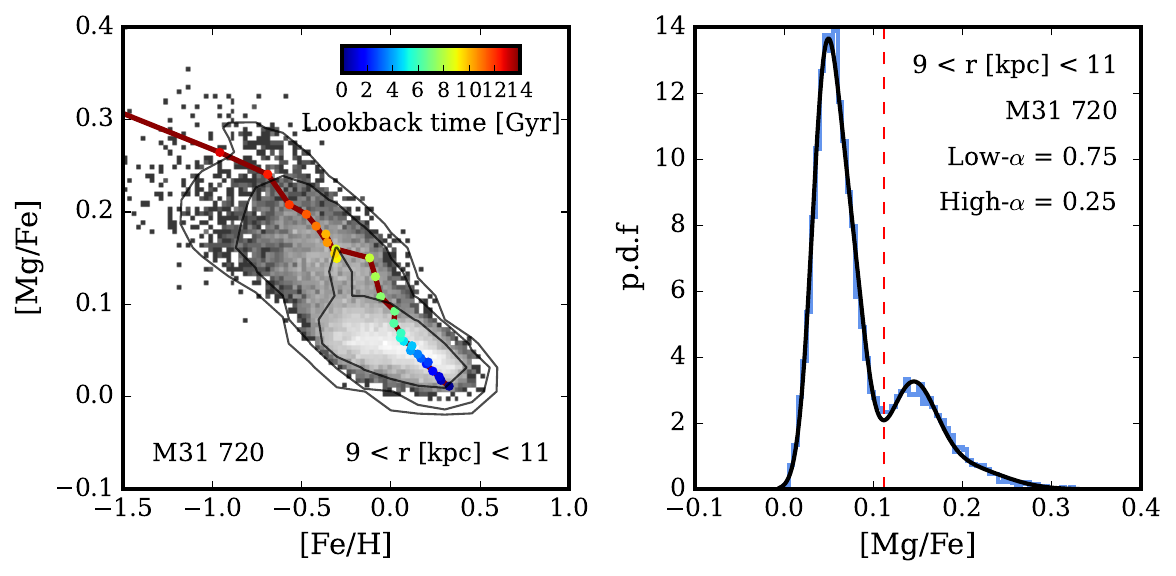}
    \caption{As Fig.~\ref{fig_8_kpc}, but for M31 720.}
    \label{figc1}
\end{figure}

\begin{figure}
    \centering
    \includegraphics[width=\columnwidth]{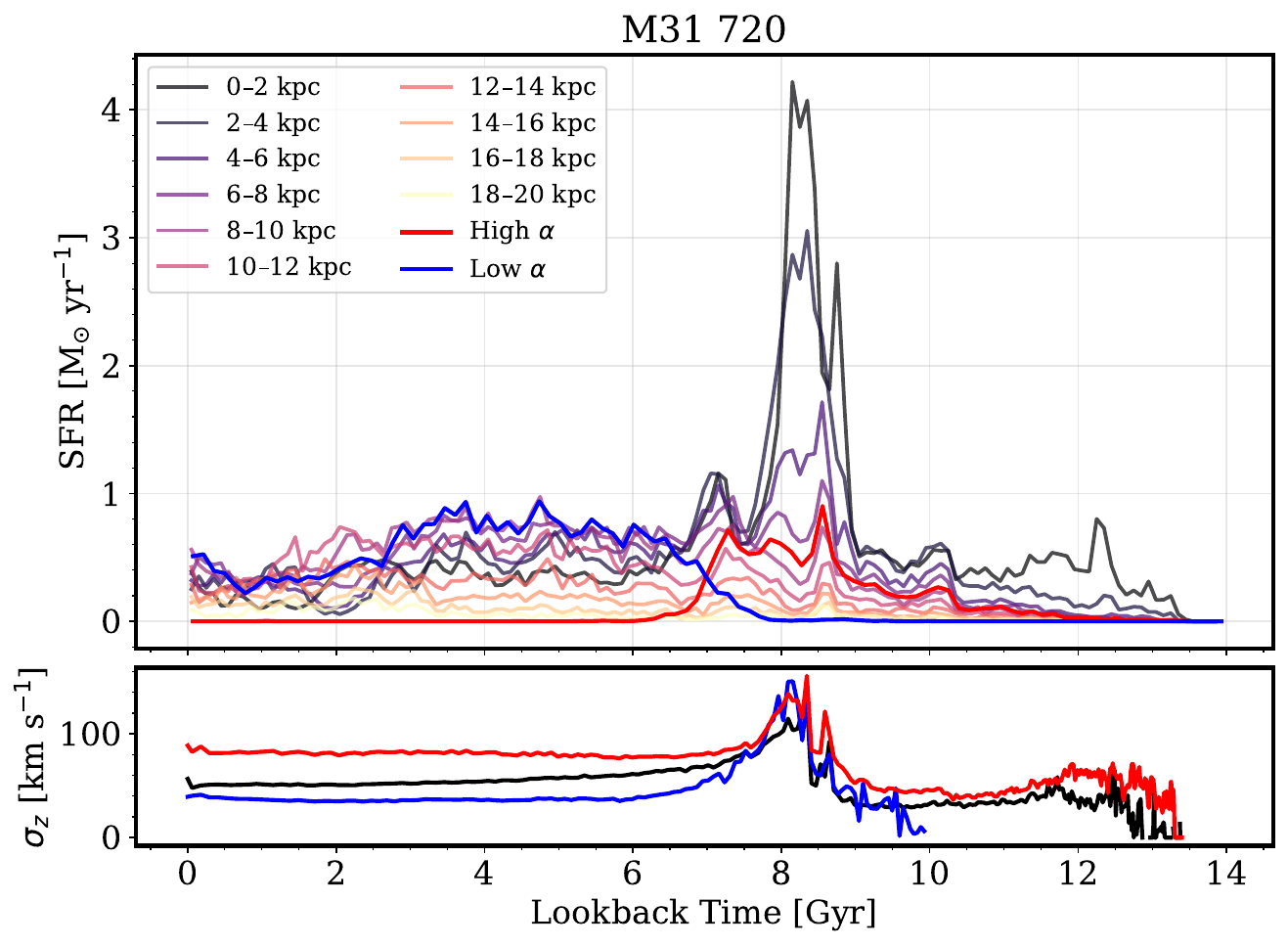}
    \caption{As Fig.~\ref{fig:sfr_disk}, but showing the star formation rate for different radii in M31 720. }
    \label{figc2}
\end{figure}


\bsp	
\label{lastpage}
\end{document}